\documentstyle[aps,prb,epsf]{revtex}  
\begin{document}

\title{ Solution of effective Hamiltonian of impurity hopping between two
         sites in a metal }
\author{Jinwu  Ye}
\address{ Physics Laboratories,       
Harvard  University, Cambridge, MA, 02138 and  \\
 Department of Physics and Astronomy, Johns Hopkins University,
 Baltimore, MD, 21218}
\date{\today}
\maketitle
\begin{abstract}

    We analyze in detail all the possible fixed points of the effective
  Hamiltonian of a non-magnetic impurity hopping between two sites in a metal
  obtained by Moustakas and Fisher(MF).  We find a 
  {\em line of non-fermi liquid fixed points } which continuously interpolates
  between the 2-channel Kondo fixed point(2CK) and the one channel, two
  impurity Kondo (2IK) fixed point. There is one relevant direction
  with scaling dimension 1/2 and one leading irrelevant operator with dimension 3/2.
   There is also one marginal operator in the {\em spin} sector moving along this line.
   The marginal operator, combined with the leading irrelevant operator,
  will generate the relevant operator.  For the general position
  on this line, the leading low temperature {\em exponents}
  of the specific heat, the {\em hopping } susceptibility and the electron conductivity 
  $ C_{imp}, \chi^{h}_{imp}, \sigma(T) $ are the same as those of the 2CK, but
  the finite size spectrum depend on the position on
  the line. {\em No} universal ratios can be formed from the amplitudes 
  of the three quantities except at the 2CK point on this line where
  the universal ratios can be formed. At the 2IK point on this line,
   $ \sigma(T) \sim 2\sigma_{u}(1 + aT^{3/2}) $, {\em no} universal ratio can be formed either.
  The additional {\em non-fermi liquid fixed point} found by MF has the
  same symmetry as the 2IK, it has {\em two} relevant directions with scaling
  dimension 1/2, therefore also unstable. The leading low temperature behaviors 
  are $ C_{imp} \sim T, \chi^{h}_{imp} \sim \log T, \sigma(T) \sim 2\sigma_{u}
  (1+ a T^{3/2} ) $, no universal ratios can be formed. The system is shown to flow
  to a {\em line of fermi-liquid fixed points } which continuously interpolates
  between the non-interacting fixed point and the 2 channel spin-flavor Kondo
  fixed point (2CSFK) discussed by the author previously.
  The effect of particle-hole symmetry breaking is discussed.
  The effective Hamiltonian in the external magnetic field is analysed.
  The scaling functions for the physical measurable quantities are
  derived in the different regimes; their predictions for the experiments
  are given. Finally the implications are given for a non-magnetic impurity hopping
  around three sites with triangular symmetry discussed by MF.

\end{abstract}
%\narrowtext
\section{Introduction}
  The experimental realization of  
  overscreened multichannel Kondo model has been vigorously searched
  since the discovery of its non-fermi liquid behavior (NFL) by
  Nozi\'{e}res and Blandin (NB) \cite{blandin}.
  D. L. Cox \cite{cox} pointed out that the NFL behaviors
  in heavy fermion systems like $ Y_{1-x} U_{x} Pd_{3} $ may be explained
  by the 2 channel quadrupolar Kondo effects, but the observed electrical
  conductivity  of such systems is linear in $ T $ in contrast to $ \sqrt{T} $
  behavior of the 2-channel Kondo model (2CK). 
  Vlad\'{a}r and Zawadowski \cite{zaw}
  suggested that a non-magnetic impurity tunneling
  between two sites in a metal can be mapped to the 2CK in which
  the roles of channels and spins in the original formulation are interchanged.
  Ralph {\sl et al.} \cite{ralph}
  proposed that the conductance signals observed in ballistic
  metal point contacts may be due to the 2-channel Kondo scattering from 2-level
  tunneling systems, the conductance exponent $ 1/2 $ and the magnetic
  field dependence observed in such device
  is indeed in consistent with that predicted by Affleck-Ludwig's (AL) Conformal
  Field Theory (CFT) solution of the 2CK \cite{affleck1,affleck2}
  however the alternative interpretation was also proposed \cite{wingreen}.

     Moustakas and Fisher \cite{fisher1} reexamined the problem of the electron assisted
  tunneling of a heavy particle between two sites in a metal. In addition to bare hopping
  ( $ \Delta_{0} $ term in Eq. \ref{start0} ) and one electron assisted hopping term
  ( $ \Delta_{1} $ term in Eq. \ref{start0} ) found previously in Ref.\cite{zaw},
  they found that an {\em extra}
  two electrons assisted hopping term (  $ \Delta_{2} $ term in Eq. \ref{start0} )
  also plays an important role. Treating all these
  important processes carefully, they concluded that more than four channels (including spin)
  are needed in order to localize the impurity. In Ref. \cite{fisher2}, they wrote down
  an effective Hamiltonian which includes all the important processes and
  employed Emery-Kivelson (EK)'s Abelian Bosonization solution of the 2CK
  \cite{emery} to investigate the full phase diagram of this Hamiltonian, they found 
  the two electron assisted hopping term plays a {\em similar} role to the bare hopping term.
  However, they overlooked the important fact
  that the canonical transformation
  operator $ U=e^{i S^{z} \Phi_{s} } $ in EK's solution is a boundary condition
  changing operator \cite{xray}, therefore
  their analysis of the symmetry of the fixed points and the operator contents
  near these fixed points
  are not complete. They didn't calculate the electron conductivity  which is
  the most important experimental measurable quantity.
  Furthermore the nature of the stable Fermi liquid
  fixed point was also left unexploited.

  Affleck and Ludwig (AL) ~\cite{bound1,xray,ludwig},
  using Conformal Field Theory, pointed out that for {\em any general} quantum impurity
  problem, the impurity degree of freedoms completely {\em disappear} from the description
  of the low temperature fixed point and leave behind  conformally 
  invariant {\em boundary conditions}. CFT can also be used to classify all the possible 
  boundary operators near any low temperature fixed points and calculate any correlation
  functions.  For 4 pieces of bulk fermions which correspond to 8 pieces of
 Majorana fermions, the non-interacting theory
 possesses $ SO(8) $ symmetry,
 Maldacena and Ludwig (ML) \cite{ludwig} showed that finding the symmetry
 of the fixed points is exactly equivalent
 to finding the boundary conditions of the 8 Majorana fermions
 at the fixed points, the boundary conditions turned out to be {\em linear}
 in the basis which separates charge, spin and flavor. 
 ML reduced the descriptions of the fixed points as free chiral bosons plus
 different {\em linear} boundary conditions.
 The linear boundary
 conditions can also be transformed into the boundary conditions in the original
 fermion basis by the triality transformation Eq.\ref{second} \cite{witten}.
 The boundary conditions
 in the original fermion basis only fall into
 two classes: {\em NFL}
 fixed points where the original fermions are scattered into spinors;
 {\em fermi liquid}(FL) fixed points where the original fermions only suffer
 just phase shifts at the boundary. 

   The important step in the CFT approach developed by AL is the identification
 of the fusion rules at various fixed points. Although the fusion rule is simple
 in the multichannel Kondo model, it is usually very difficult to identify in more
 complicated models like the one discussed in this paper.

Recently, using EK's Abelian Bosonization approach
 to the 2CK, the author developed a simple
 and powerful method to study certain class of quantum impurity models
 with 4 pieces of bulk fermions. the method can identify very quickly
 all the possible boundary fixed points and their maximum symmetry,
 therefore circumvent the difficult tasks to identify the fusion rules at
 different fixed point or line of fixed points.
 it can also demonstrate the physical picture at the boundary explicitly \cite{powerful}.

  In this paper, using the method developed in Ref.\cite{powerful} and paying the special
  attention to boundary condition changing nature of $ U=e^{i S^{z} \Phi_{s} } $,
  we investigate the full phase diagram of the present problem again.
 In Sec. II, we Abelian bosonize the effective Hamiltonian. By using the
  Operator Product Expansion (OPE) \cite{cardy}, we get the 
  Renormalization Group(RG) flow equations near the weak coupling line of fixed points ,
  therefore identify the two {\em independent} crossover scales. In the following
  sections, we analyze all the possible fixed points of the bosonized Hamiltonian.  
 In Sec.III, we find  {\em a line of NFL fixed points} which continuously interpolates
  between the 2CK fixed point and the one channel two impurity
  Kondo (2IK) fixed point \cite{twoimp}, its symmetry is
   $ U(1) \times O(1) \times O(5) $ \cite{ising}. This  line of NFL fixed points  is unstable,
   it has {\em one} relevant direction with scaling dimension 1/2. It also has one marginal
   operator in the spin sector which is responsible for this line. The OPE of this marginal operator
   and the leading irrelevant operator will always generate the relevant operator.
    For the general position on the line, although the leading exponents of
   the specific heat, {\em hopping } susceptibility
   and the electron conductivity  $ C_{imp}, \chi^{h}_{imp}, \sigma(T) $ are the same as those of the 2CK,
   the finite size spectrum depend on the position on the line;
   no universal relations can be found among the amplitudes of the
   three quantities. Only at the 2CK point on the line, universal ratios can be formed.
   However, at the 2IK point, the coefficient of $ \sqrt{T} $ vanishes;
   we find two dimension 5/2 operators which lead to
    $ \sigma(T) \sim 2 \sigma_{u}(1 + T^{3/2}) $,
  {\em no} universal ratio can be formed either.
  In Sec.IV, the additional {\em NFL fixed point} found by MS is shown to have the symmetry
  $ O(7) \times O(1) $, therefore is the same fixed point as the 2IK. This fixed point
  is also unstable, it has {\em two} relevant directions with scaling dimension 1/2.
  Because the leading irrelevant operators near this NFL fixed point
  are {\em first order Virasoro descendant } with scaling dimension 3/2 which
  can be written as a total imaginary time derivative, therefore, can be droped.
   The subleading irrelevant operators with dimension 2 give $ C_{imp} \sim T $.
   However, because the 'orbital field' in Eq \ref{start}
  couples to a {\em non-conserved } current,  $ \chi^{h}_{imp} \sim \log T $.
  We also find two dimension 5/2 irrelevant operators, one of them contributes to the leading
  low temperature conductivity  $ \sigma(T) \sim 2 \sigma_{u}(1+ T^{3/2}) $. No universal
  ratios can be formed near this NFL fixed point.
  In Sec. V, we find the system flows to a stable {\em line of FL fixed points}
  which continuously interpolates between the non-interacting
  fixed point and the 2 channel spin-flavor Kondo (2CSFK) fixed point discussed by the
  author in Ref. \cite{spinflavor}, its symmetry is $ U(1) \times O(6) \sim U(4) $.
  Along this line of fixed points , the electron fields of the even and
   the odd parity under interchanging the two sites suffer opposite continuously- changing
   phase shifts. We also discuss the effect of the marginal operator in the charge sector
   due to the P-H symmetry breaking and compare it with the magical operator 
   in the spin sector. In Sec. VI, we analyse the effective Hamiltonian in
    the external magnetic field which break the channel symmetry. In Sec. VII,
   all the scaling functions for the physical measurable quantities including
   the real spin susceptibility are derived in different regimes.
   In the final section, the relevance
   of the results of this paper to the experiments are examined,
   some implications on the non-magnetic impurity hopping around three sites with
   triangular symmetry are also given.
   In Appendix A, the finite size spectrum of one {\em complex} fermion is listed.
   In Appendix B, the boundary conditions in the original fermion basis are derived
   by both Bosonization method and $\gamma $ matrix method. 
   In Appendix C, the results on the additional NFL fixed point in Sec. IV. are 
   rederived in a different basis.

\section{ Bosonization of the effective Hamiltonian and the weak coupling analysis}
   We start from the following effective Hamiltonian for a non-magnetic
   impurity hopping between two sites
   in a metal first obtained by MF \cite{fisher1}, \cite{fisher2} :
 \begin{eqnarray}
 H &= & H_{0}
  + V_{1} (\psi^{\dagger}_{1 \sigma} \psi_{1 \sigma}+
    \psi^{\dagger}_{2 \sigma} \psi_{2 \sigma})   \nonumber   \\
   & + & V_{2} (\psi^{\dagger}_{1 \sigma} \psi_{2 \sigma}+
    \psi^{\dagger}_{2 \sigma} \psi_{1 \sigma} )
  + V_{3} ( d^{\dagger}_{1} d_{1}- d^{\dagger}_{2} d_{2} )
   (\psi^{\dagger}_{1 \sigma} \psi_{1 \sigma}- \psi^{\dagger}_{2 \sigma}
    \psi_{2 \sigma})
                                       \nonumber \\
  & + & d^{\dagger}_{2} d_{1} ( \frac{ \Delta_{0} }{2 \pi \tau_{c} } 
      +\frac{ \Delta_{1} }{2} \sum_{\sigma} \psi^{\dagger}_{1 \sigma}
     \psi_{2 \sigma}+ \Delta_{2} 2 \pi \tau_{c} 
  \psi^{\dagger}_{1 \uparrow} \psi_{2 \uparrow}
   \psi^{\dagger}_{1 \downarrow} \psi_{2 \downarrow} ) +h.c. + \cdots
\label{start0}
 \end{eqnarray}

   Here the two sites $ 1,2 $ ( the two real spin directions $\uparrow, \downarrow $ )
   play the role of the two spin directions $ \uparrow, \downarrow $
   ( the two channels $ 1, 2 $ ) in the
   magnetic Kondo model. All the couplings have been made to be {\em dimensionless}.
   As emphasized by MF, even initially $ \Delta_{1}, \Delta_{2} $
   maybe negligible, they will be generated at lower energy scales, $ \cdots $
   stands for irrelevant terms \cite{note}.  In the following, we use the notation of 
   the magnetic Kondo model and rewrite the above Hamiltonian as:

 \begin{eqnarray}
 H &= & H_{0} + V_{1} J_{c} (0) + 2 V_{2} J^{x}(0) +
   4 V_{3} S^{z} J^{z}(0) +\Delta_{1}  (J^{x}(0) S^{x} + J^{y}(0) S^{y} )
   \nonumber   \\
  & + & \frac{ \Delta_{0} }{ \pi \tau_{c} } S^{x}
      +\Delta_{2} 2 \pi \tau_{c} ( S^{-}
  \psi^{\dagger}_{1 \uparrow} \psi_{1 \downarrow}
   \psi^{\dagger}_{2 \uparrow} \psi_{2 \downarrow}  +h.c.) + h(\int dx J^{z}(x) + S^{z})
\label{start}
 \end{eqnarray}

  where $ S^{+} =d_{1}^{\dagger} d_{2}, S^{-} =d_{2}^{\dagger} d_{1},
      S^{z} =\frac{1}{2} ( d_{1}^{\dagger} d_{1} - d_{2}^{\dagger} d_{2} ) $.
     We also add a 'uniform  field' which corresponds to strain or pressure
     in the real experiment of Ref.\cite{ralph}. The real magnetic field
     will break the channel symmetry, its effect will be discussed in Sec. VI. 

  Hamiltonian Eq. \ref{start} has the global $ Z_{2} \times
   SU_{f}(2) \times U_{c}(1) $  symmetry and Time reversal symmetry.
 Under the $ Z_{2} $ symmetry in the spin sector : $ \psi_{i \uparrow}
   \longleftrightarrow \psi_{i \downarrow } ,
  S^{x} \rightarrow S^{x}, S^{y} \rightarrow - S^{y}, S^{z}
   \rightarrow -S^{z} $.
 Under Time reversal symmetry \cite{atten} : $ i \rightarrow -i,
    \psi_{L} \rightarrow  \psi_{R},
  S^{x} \rightarrow S^{x}, S^{y} \rightarrow - S^{y},
  S^{z} \rightarrow S^{z} $. The potential scattering term $ V_{1} $ is 
  the {\em  only } term which breaks P-H symmetry:  
  $\psi_{i \alpha}(x) \rightarrow  \epsilon_{\alpha \beta}
  \psi^{\dagger}_{j \beta}(x) $. In contrast to the 2CK, we only have $ Z_{2} $ symmetry
 in the spin sector, the total spin current to  which $ h $ couples is not conserved

       In the following, we closely follow the notations of Emery-Kivelson
 \cite{emery}. Abelian-bosonizing the four bulk Dirac fermions separately :
\begin{equation}
 \psi_{i \alpha }(x )= \frac{ P_{i \alpha}}{\sqrt{ 2 \pi \tau_{c} }}
  e ^{- i \Phi_{i \alpha}(x) }
\label{first}
\end{equation}
    Where  $ \Phi_{i \alpha} (x) $ are the real chiral bosons satisfying
  the commutation relations
\begin{equation}
  [ \Phi_{i \alpha} (x), \Phi_{j \beta} (y) ]
   =    \delta_{i j} \delta_{\alpha \beta} i \pi sgn( x-y )
\end{equation}

 The cocycle factors
  have been chosen as: $ P_{1 \uparrow}  =  P_{1 \downarrow } = e^{i \pi N_{1 \uparrow} }
   , P_{2 \uparrow}  =   P_{2 \downarrow } = e^{i \pi (
    N_{1 \uparrow}  + N_{ 1 \downarrow} +N_{2 \uparrow} )} $.

   It is convenient to introduce the following charge, spin, flavor,
  spin-flavor bosons:
\begin{equation}
   \left ( \begin{array}{c} \Phi_{c} \\
 \Phi_{s}  \\
 \Phi_{f}  \\
 \Phi_{sf} \end{array}  \right )
  =\frac{1}{2} \left ( \begin{array}{c} \Phi_{1 \uparrow }+ \Phi_{1\downarrow }+
            \Phi_{2 \uparrow }+ \Phi_{2 \downarrow }  \\
           \Phi_{1 \uparrow }- \Phi_{1\downarrow }+
            \Phi_{2 \uparrow }- \Phi_{2 \downarrow }  \\
           \Phi_{1 \uparrow }+ \Phi_{1\downarrow }-
            \Phi_{2 \uparrow }- \Phi_{2 \downarrow }  \\
           \Phi_{1 \uparrow }- \Phi_{1\downarrow }-
            \Phi_{2 \uparrow }+ \Phi_{2 \downarrow }  \end{array} \right )
\label{second}  
\end{equation}

   Following the four standard steps in EK solution: {\em step 1 :}
  writing the Hamiltonian in terms of the chiral bosons Eq.\ref{second}.
 {\em step 2:} making the canonical transformation
    $ U= \exp ( -i V_{1} \Phi_{c} (0)+
 i S^{z} \Phi_{s}(0)) $. {\em step 3} shift the spin boson
 by $ \partial_{x} \Phi_{s} \rightarrow \partial_{x}\Phi_{s} +
\frac{h}{v_{F}} $ \cite{gogolin}.  {\em step 4:} making the following refermionization:

\begin{eqnarray}
S^{x} &= & \frac{ \widehat{a}}{\sqrt{2}} e^{i \pi N_{sf}},~~~
S^{y}= \frac{ \widehat{b}}{\sqrt{2}} e^{i \pi N_{sf} },~~~
S^{z}= -i \widehat{a} \widehat{b} = d^{\dagger} d - \frac{1}{2}        \nonumber \\
 \psi_{sf} & = & \frac{1}{\sqrt{2}}( a_{sf} - i b_{sf} ) =
  \frac{1}{\sqrt{ 2 \pi \tau_{c}}} e^{i \pi N_{sf} }
                e^{-i \Phi_{sf} }     \nonumber   \\
 \psi_{s,i} & = & \frac{1}{\sqrt{2}}( a_{s,i} - i b_{s,i} )=
 \frac{1}{\sqrt{ 2 \pi \tau_{c}}} e^{i \pi d^{\dagger} d } e^{i \pi N_{sf} }
 e^{-i \Phi_{s} }    
\label{ek}
\end{eqnarray}

  Note  $ \psi_{s,i}(x) $ defined above contains the impurity operator $ e^{ i \pi
  d^{\dagger} d } $ in order to satisfy the anti-commutation relations with the other
  fermions.

    The transformed Hamiltonian $ H^{\prime}= U H U^{-1} $ can be written
  in terms of the Majorana fermions as: 

 \begin{eqnarray}
 H^{\prime}  &= & H_{0}
  + 2 y \widehat{a} \widehat{b} a_{s,i}(0) b_{sf}(0)
  + 2 q  \widehat{a} \widehat{b} a_{s,i}(0) b_{s,i}(0)-i \widehat{a} \widehat{b} q \frac{h}{v_{F}}
                       \nonumber  \\
  & - & i \frac{ \Delta_{1}}{\sqrt{ 2 \pi \tau_{c} }} \widehat{a} b_{sf}(0)
   +i \frac{ \Delta_{+}}{\sqrt{ 2 \pi \tau_{c} }} \widehat{b} a_{s,i}(0)
   +i \frac{ \Delta_{-}}{\sqrt{ 2 \pi \tau_{c} }} \widehat{a} b_{s,i}(0)
\label{group}
 \end{eqnarray}

    where $ y=2 V_{2}, q=\frac{1}{ \pi} ( V_{3}- \frac{ \pi v_{F}}{2} )   
    ,  \Delta_{\pm} = \Delta_{0} \pm \Delta_{2} $.

 As observed by MF, the above equation clearly indicate that
  the two electron assisted hopping term plays a similar role
 to the bare hopping term.
   From the OPE \cite{cardy}
   of the various operators in Eq.~\ref{group},
   the R. G. flow equations near the weak coupling fixed point $ q=0 $ is \cite{anti} 
\begin{eqnarray}
\frac{d \Delta_{+}}{d l} & = & \frac{1}{2} \Delta_{+} + 2y \Delta_{1}   \nonumber   \\
\frac{d \Delta_{-}}{d l} & = & \frac{1}{2} \Delta_{-}   \nonumber    \\
\frac{d \Delta_{1}}{d l} & = & \frac{1}{2} \Delta_{1} + 2y \Delta_{+}  \nonumber  \\
\frac{d y}{d l} & = & \Delta_{1} \Delta_{+}   \nonumber \\
\frac{d q}{d l} & = & \Delta_{+} \Delta_{-}
\label{weak1}
\end{eqnarray}

   MF got the same RG flow equations ( Eq.23 in Ref.\cite{fisher2} )
   near the weak coupling  line of fixed points by using Anderson-Yuval Coulomb gas picture.
  Eq. \ref{weak1} shows that $ \Delta_{+}, \Delta_{-}, \Delta_{1} $ have the same
  scaling dimension 1/2 at $ q=0 $, so are equally important.

  Define $\tilde{b}_{s,i}(x), \tilde{b}_{sf,i}(x)$ (similarly for  $\tilde{ q }, \tilde{y} $) 
\begin{equation}
 \left( \begin{array}{c} \tilde{b}_{s,i}(x)   \\
                         \tilde{b}_{sf,i}(x)   \end{array}  \right )
   = \frac{1}{ \Delta_{K} } \left( \begin{array}{cc}
        \Delta_{1}    &   \Delta_{-}   \\
       - \Delta_{-}    &   \Delta_{1}   \\
       \end{array}  \right )
 \left( \begin{array}{c} b_{s,i}(x)   \\
                         b_{sf}(x)   \end{array}  \right )
\label{twist}
\end{equation}

  Where $ \Delta_{K} =\sqrt{ \Delta_{1}^{2} +\Delta_{-}^{2} }  $.

  Eq.\ref{group} can be rewritten as:
\begin{eqnarray}
 H^{\prime}   =    H_{0}
  +  2 \tilde{q} \widehat{a} \widehat{b} a_{s,i}(0) \tilde{b}_{s,i}(0)
  + 2 \tilde{y} \widehat{a} \widehat{b} a_{s,i}(0) \tilde{b}_{sf,i}(0) \nonumber  \\
    -    i \frac{ \Delta_{K}}{\sqrt{ 2 \pi \tau_{c} }}
      \widehat{a} \tilde{b}_{sf,i}(0)
   +i \frac{ \Delta_{+}}{\sqrt{ 2 \pi \tau_{c} }} \widehat{b} a_{s,i}(0)
   -i \widehat{a}\widehat{b} q
     \frac{h}{v_{F}}
\label{final}
\end{eqnarray}
   
    The R. G. flow equations which are equivalent to Eq.\ref{weak1} are (we set $ h=0 $ ) 
\begin{eqnarray}
\frac{d \Delta_{+}}{d l} & = & \frac{1}{2} \Delta_{+} + 2 \tilde{y} \Delta_{1}   \nonumber   \\
\frac{d \Delta_{K}}{d l} & = & \frac{1}{2} \Delta_{K} + 2 \tilde{y} \Delta_{+}   \nonumber    \\
\frac{d \theta }{d l} & = & 4 \tilde{q} \frac{\Delta_{+}}{\Delta_{K}}  \nonumber  \\
\frac{d \tilde{y}}{d l} & = & \Delta_{K} \Delta_{+}   \nonumber \\
\frac{d \tilde{q}}{d l} & = & 0
\label{weak2}
\end{eqnarray}

    Where the angle $\theta $ is defined by $ \cos\theta =\frac { \Delta_{-}^{2}
         - \Delta_{1}^{2} }{ \Delta_{K}^{2}},
     \sin \theta =\frac { 2 \Delta_{-} \Delta_{1} }{ \Delta_{K}^{2}} $.

  The crossover scale from the weak coupling fixed point $ q=0 $ to 
  a given point on the  line of NFL fixed points  to be discussed in section III is given by
  $ T_{K1} \sim D ( \Delta_{K} )^{2} $, to the additional NFL fixed point 
  to be discussed in section IV.
  is given by $ T_{K2} \sim D (\Delta_{+})^{2} $.

  As emphasized in \cite{xray,powerful}, the canonical transformation  $ U $ is
 a boundary condition changing operator,
 the transformed field $ \psi^{\prime}_{s}(x) $ is related to the original
  field $ \psi_{s}(x) $ by \cite{not}:
\begin{equation}
\psi^{\prime}_{s}(x) = U^{-1} \psi_{s,i}(x) U =
   e^{i \pi d^{\dagger} d } e^{i \pi S^{z} sgnx} \psi_{s}(x)=-isgnx \psi_{s}(x)
\label{define}
\end{equation}
   As expected, the impurity spin $ S^{z} $ {\em drops } out in the prefactor of
  the above equation.

    The above Eq. can be written out explicitly in terms of the Majorana fermions
\begin{eqnarray}
    a^{\prime L}_{s}(0)=- b^{L}_{s}(0),~~~~ b^{\prime L}_{s}(0)= a^{L}_{s}(0)  \nonumber  \\
    a^{\prime R}_{s}(0)= b^{R}_{s}(0),~~~~ b^{\prime R}_{s}(0)=- a^{R}_{s}(0)
\label{general}
\end{eqnarray}

  We find the physical picture can be more easily demonstrated in
  the corresponding action:

\begin{eqnarray}
 S &= & S_{0} +  \frac{\gamma_{1}}{2} \int d\tau \widehat{ a}(\tau)
     \frac{ \partial \widehat{a}(\tau)}{\partial \tau}
   -  i \frac{ \Delta_{K}}{\sqrt{ 2 \pi \tau_{c} }} \int d \tau
      \widehat{a}(\tau) \tilde{b}_{sf,i}(0, \tau)
                              \nonumber   \\
  &   +  &   \frac{\gamma_{2}}{2} \int d\tau  \widehat{b}(\tau)
   \frac{ \partial \widehat{b}(\tau)}{\partial \tau}
   +i \frac{ \Delta_{+}}{\sqrt{ 2 \pi \tau_{c} }} \int d \tau
    \widehat{b}(\tau) a_{s,i}(0, \tau)
                                      \nonumber  \\
 & + & 2 \tilde{q} \int d \tau \widehat{a}(\tau) \widehat{b}(\tau)
    a_{s,i}(0, \tau) \tilde{b}_{s,i}(0,\tau)
  + 2 \tilde{y} \int  d \tau  \widehat{a}(\tau) \widehat{b}(\tau)
    a_{s,i}(0, \tau) \tilde{b}_{sf,i}(0, \tau)
\label{action}
\end{eqnarray}

   When performing the RG analysis of the action $ S $, we keep \cite{known}
  1: $\gamma_{2}=1 $, $ \Delta_{K}$ fixed,
  2: $\gamma_{1}=1 $, $ \Delta_{+}$ fixed,
  3: $\Delta_{K}, \Delta_{+}$ fixed.

   We will identify all the possible fixed points  or the line of fixed
  points and derive the R. G. flow equations near these fixed points
  in the following sections respectively.

\section{ The line of NFL fixed points }
 If $ \tilde{q}=\tilde{y} =\Delta_{+} =0 $,
 this fixed point is located
 at $ \gamma_{1}=0, \gamma_{2}=1 $ where $\widehat{b} $ decouples,
 but $ \widehat{a} $ loses its kinetic energy and becomes a
 Grassmann Lagrangian multiplier, integrating $ \widehat{a} $ out leads to
 the {\em boundary conditions} :

\begin{equation}
   \tilde{b}^{L}_{sf,i}(0)= -\tilde{b}^{R}_{sf,i}(0) 
\label{bound1p}
\end{equation}
   
   We also have the trivial boundary conditions
\begin{equation}
 a^{L}_{s,i}(0)=a^{R}_{s,i}(0) ,~~ \tilde{b}^{L}_{s,i}(0)= \tilde{b}^{R}_{s,i}(0) 
\label{bound2p}
\end{equation}

  By using Eqs.\ref{twist},\ref{define}, we find the boundary conditions in $ H^{\prime} $ \cite{scaling}
\begin{equation}
 a^{\prime L}_{s}(0)  =  a^{\prime R}_{s}(0),~~~~
 \left( \begin{array}{c} b^{\prime L}_{s}(0)   \\
                         b^{\prime L}_{sf}(0)   \end{array}  \right )
   =  \left( \begin{array}{cc}
        -\cos \theta    &   \sin \theta   \\
       \sin \theta    &   \cos \theta   \\
       \end{array}  \right )
 \left( \begin{array}{c} b^{\prime R}_{s}(0)   \\
                         b^{\prime R}_{sf}(0)   \end{array}  \right )
\end{equation}

   By using Eq.\ref{general}, we get the corresponding boundary conditions in $ H $
\begin{equation}
 b^{L}_{s}(0)  =  -b^{R}_{s}(0) ,~~~ \psi^{L}_{a}(0) =  e^{i 2 \delta} \psi^{R}_{a}(0),~~~
      \theta=2 \delta 
\label{vectornf}
\end{equation}
    where $ \psi_{a}(0) = a_{s}(0) - i b_{sf}(0)$.

  It is evident that at the fixed point, the impurity degree of freedoms
 totally {\em disappear} and leave behind the conformally invariant boundary
 conditions Eq.\ref{vectornf}.
  These are  {\em NFL} boundary conditions. {\em In contrast to} the boundary
  conditions discussed previously ~\cite{ludwig,powerful,flavor,spinflavor},
  although the boundary
  conditions are still linear in the basis which separates charge, spin  and flavor,
  they are {\em not } in any four of the Cartan subalgebras of $ SO(8) $ group \cite{witten},
  therefore {\em cannot} be expressed in terms of the chiral bosons in Eq.\ref{second}.

  However,  Eq.\ref{vectornf} indicates that 
  it is more convenient to regroup the  Majorana fermions as \cite{decouple,cut}
\begin{eqnarray}
 \psi^{n}_{s} & = & a_{s}-i b_{sf}= e^{-i \Phi^{n}_{s}}   \nonumber   \\
 \psi^{n}_{sf} & =  & a_{sf}+i b_{s}= e^{-i \Phi^{n}_{sf}} 
\label{jinwu}
\end{eqnarray}

  The boundary condition Eq.\ref{vectornf} can be expressed in terms of
  the {\em new} bosons
\begin{equation}
 \Phi^{n}_{s,L}=\Phi^{n}_{s,R}+ \theta,~~~~ \Phi^{n}_{sf,L}= - \Phi^{n}_{sf,R}
\label{obi}
\end{equation}

\begin{figure}
\epsfxsize=10 cm
\centerline{\epsffile{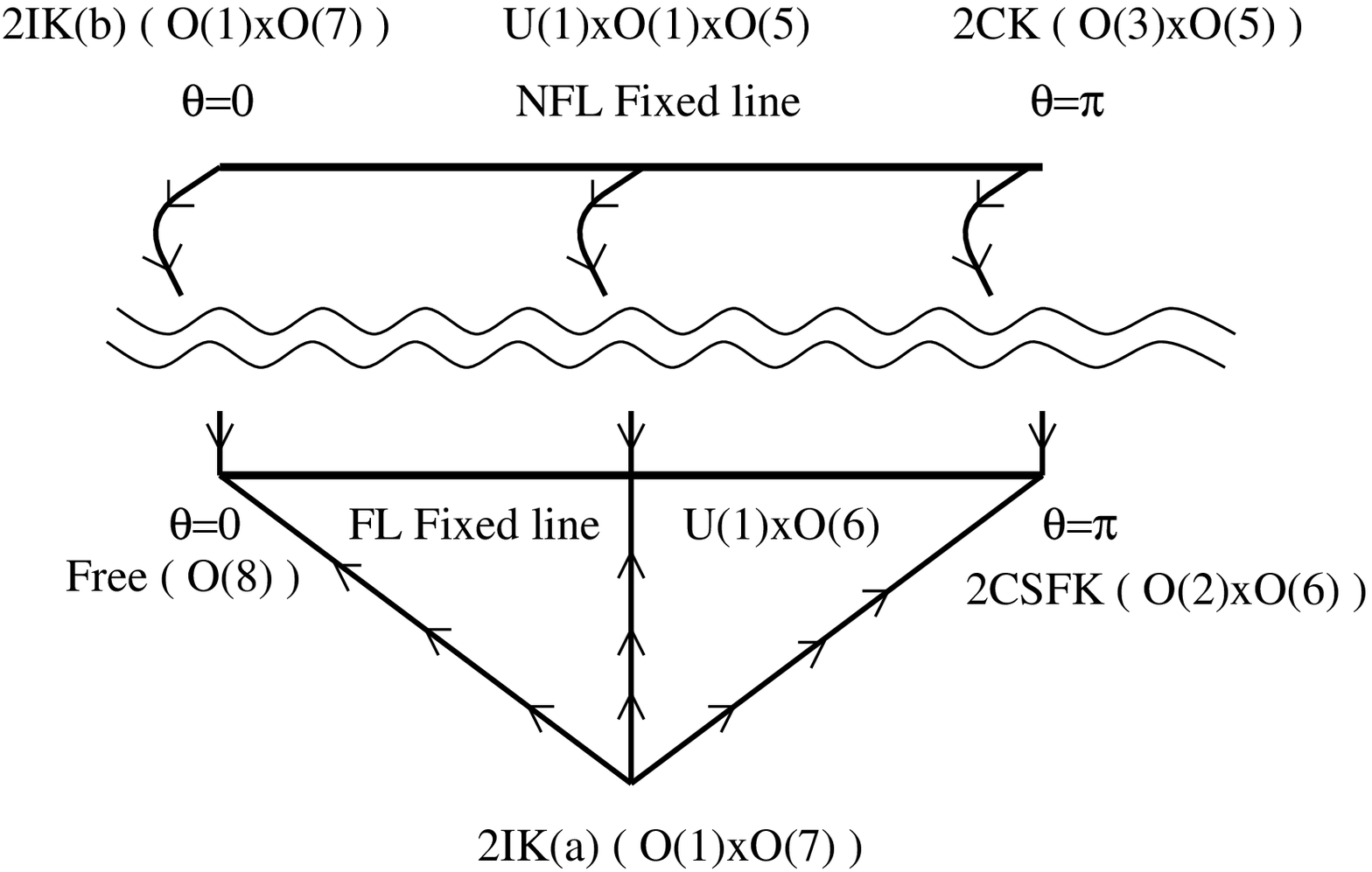}}
\caption{ Phase diagram of a non magnetic impurity hopping between two sites
   in a metal.  The {\em  line of NFL fixed points  } has the  symmetry $ O(1) \times U(1) \times O(5) $
   with $ g=\sqrt{2} $ which interpolates continuously between
   the 2CK fixed point and the 2IK fixed point.
   If $ \theta=\pi $, the fixed point
   symmetry is enlarged to $ O(3) \times O(5) $ which is
   the fixed point symmetry of the 2CK. If $ \theta=0 $,  the fixed point
   symmetry is enlarged to $ O(1) \times O(7) $ which is
   the fixed point symmetry of the 2IK(b). This {\em  line of NFL fixed points } is unstable,
   there is {\em one} relevant operator with dimension 1/2 which drives the system
   to the {\em  line of FL fixed points }. There is also a marginal operator
   along the line.
   This {\em  line of FL fixed points  } has the symmetry $ U(1) \times O(6) \sim U(4) $ with $ g=1 $ 
   which interpolates continuously between the non-interacting
   fixed point and the 2CSFK fixed point.
   If $ \theta=0 $, the fixed point
   symmetry is enlarged to $ O(8) $ which is the fixed point symmetry of the
   non-interacting electrons.  If $ \theta=\pi $, the fixed point
   symmetry is enlarged to $ O(2) \times O(6) $ which is the the fixed point
   symmetry of the 2CSFK. This {\em  line of FL fixed points } is stable.
    There is a marginal operator along this line.
    The additional {\em NFL fixed point } with $ g=\sqrt{2} $ has the symmetry $ O(1) \times O(7) $
   which is the fixed point symmetry of the 2IK(a).
   This additional NFL fixed point is also {\em unstable}.
   There are {\em two} relevant terms with scaling
   dimension 1/2, any linear combination of the two terms will
   drive the system to a given point of the {\em  line of FL fixed points }.
   See the text for the detailed discussions on the physical properties
   of these fixed points or line of fixed points s and the crossovers between them.}
\label{picture}
\end{figure}

  As pointed out in Ref. \cite{ludwig}, in the basis of Eq.\ref{second},
  the {\em physical fermion fields} transform
  as the spinor representation of $ SO(8) $, therefore in order to find the corresponding
  boundary conditions in the physical fermion basis, we have to find the $ 16 \times 16 $
  dimensional spinor representation of boundary conditions Eq. \ref{vectornf}.
  The derivation of the boundary conditions is given in Appendix B,
  the results are found to be :
\begin{equation}
 \psi^{L}_{i \pm }  =   e^{  \pm i \theta/2 } S^{R}_{i \pm}, ~~~~~
 S^{L}_{i \pm }  =   e^{  \pm i \theta/2 } \psi^{R}_{i \pm}
\label{spinor1}
\end{equation}
   Where the new fields are defined by :
\begin{equation}
 \psi_{i \pm }   =    \frac{1}{\sqrt{2}}
      ( \psi_{i \uparrow} \pm  \psi_{i \downarrow}  ) , ~~~~~
 S_{i \pm }   =  \frac{1}{\sqrt{2}}
      ( S_{i \uparrow} \pm  S_{i \downarrow}  )  
\end{equation}
    It can be checked explicitly the above boundary conditions satisfy all the symmetry
    requirement( namely $ Z_{2} \times SU_{f}(2) \times U_{c}(1) $, Time Reversal and 
    P-H symmetry).
   Fermions with the even and the odd parity under $ Z_{2} $ symmetry are scattered into the
   collective excitations of the corresponding parity which fit into
   the $ S $ spinor representation
   of $ SO(8) $, therefore the one particle S matrix and the residual electrical
   conductivity  are the {\em same}
   with those of the 2CK.  This is a {\em  line of NFL fixed points  } with $ g=\sqrt{2} $
   and the  symmetry $ O(1) \times U(1) \times O(5) $ which interpolates continuously between
   the 2CK fixed point and the 2IK fixed point.
    If $ \theta=\pi $, namely $ \Delta_{-}=0 $, the fixed point
    symmetry is enlarged to $ O(3) \times O(5) $ which is
   the fixed point symmetry of the 2CK. If $ \theta=0 $, namely $ \Delta_{1}=0 $, the fixed point
    symmetry is enlarged to $ O(1) \times O(7) $ which is
    the fixed point symmetry of the 2IK(b) (Fig.\ref{picture}).
   The finite size $ -l < x < l $ spectrum
    ( in terms of unit $ \frac{ \pi v_{F} }{ l} $ )
   can be obtained from the free fermion spectrum with {\em both}
   periodic (R) {\em and} anti-periodic (NS) boundary conditions by twisting the
   three Majorana fermions in the spin sector:
   changing the boundary condition of the Majorana fermion $ b_{s} $ from NS sector
   to R sector or vice versa, twisting the complex fermion $\psi_{s} $ by a
   continuous angle $\theta =2 \delta $.
   The finite size spectrum of one complex fermion is derived in Appendix A.
    The complete finite size spectrum of this NFL fix line
    is listed in Table \ref{nflpositive} if $ 0< \delta < \frac{\pi}{2} $ and
    in Table \ref{nflnegative} if $ -\frac{\pi}{2} < \delta < 0 $.
    The ground state energy is
    $ E_{0} = \frac{1}{16} + \frac{1}{2} (\frac{\delta}{\pi} )^{2} $ with degeneracy $d=2$,
    the first excited energy is
    $ E_{1}-E_{0} = \frac{3}{8} - \frac{ |\delta| }{2\pi} $ ( if $ -\frac{\pi}{4}
    < \delta < \frac{\pi}{2} $ ) or
    $ E_{1} -E_{0} = \frac{1}{2} + \frac{\delta} {\pi} $ 
    (if $ -\frac{\pi}{2} < \delta < -\frac{\pi}{4} $) with $ d=4 $.

    If $ \delta=\pm \frac{\pi}{2} ( \delta=0 ) $,
    the finite size spectrum of the 2CK ( the 2IK) is recovered.
    The finite size spectrum of the 2CK and the 2IK are listed in Tables \ref{2CK}
    and \ref{2IK} respectively.

  The local correlation functions at this  line of NFL fixed points  are \cite{powerful}
\begin{equation}
\langle \widehat{a}(\tau) \widehat{a}(0) \rangle = \frac{1}{\tau},~~~~
\langle \tilde{b}_{sf,i}(0, \tau) \tilde{b}_{sf,i}(0,0) \rangle = \frac{\gamma^{2}_{1}}{\tau^{3}}
\label{local1}
\end{equation}

 We can also read the scaling dimension of the various fields
  $ [\widehat{b}]=0, [\widehat{a}]=1/2, [a_{s,i}]=[\tilde{b}_{s,i}]=1/2,
  [\tilde{b}_{sf,i}]=3/2 $.

   As shown in Ref.\cite{powerful}, at the line of fixed points , the impurity degree of freedoms
  completely disappear: $ \widehat{b} $ decouples and $ \widehat{a} $ turns into
  the {\em non-interacting} scaling field at the fixed point
\begin{equation}
      \widehat{a}(\tau) \sim  \tilde{b}_{sf,i}(0, \tau)
\end{equation}

    The corresponding two scaling fields in $ H $ is
\begin{equation}
      \frac{1}{\Delta_{K}}( -\Delta_{-} a_{s} + \Delta_{1} b_{sf}) 
\end{equation}

     Following Ref.\cite{flavor}, we find the impurity spin turns into
\begin{eqnarray}
  S_{x}(\tau) &\sim & - i( \widehat{b} b_{s} + 
      \frac{ \Delta_{1}}{\Delta_{K}}a_{s}b_{sf} ) +\cdots   \nonumber  \\ 
  S_{y}(\tau) & \sim & i( \widehat{b} a_{s} + 
      \frac{1}{\Delta_{K}}( -\Delta_{-} a_{s} + \Delta_{1} b_{sf})b_{s} ) +\cdots   \nonumber  \\ 
  S_{z}(\tau) & \sim & i \widehat{b}
    \frac{1}{\Delta_{K}}( -\Delta_{-} a_{s} + \Delta_{1} b_{sf}) +\cdots 
\end{eqnarray}
    Where $\cdots $ stands for higher dimension  operators \cite{flavor} and
  $ \frac{\Delta_{1}}{ \Delta_{K}} =\sqrt{ \frac{1-\cos \theta}{2}},
   \frac{\Delta_{-}}{ \Delta_{K}} =\sqrt{ \frac{1+\cos \theta}{2}} $.

  The impurity spin-spin correlation function $ \langle S^{a}(\tau) S^{a}(0) \rangle
  \sim 1/\tau $.

    From Eq.\ref{group}, it is easy to see
   that even $ y $ term itself is irrelevant near this line of fixed points ,
   but  Eq. \ref{weak1} shows that
   it, when combined with $ \Delta_{0} $ term, will generate
  $ \Delta_{1}, \Delta_{2} $ terms which
    must be taken into account at this line of fixed points . $ y $ term is " dangerous"
  irrelevant, similar " dangerous" irrelevant term occurred
   in the two channel flavor anisotropic Kondo model \cite{flavor}.
   This  line of NFL fixed points  is {\em unstable}.
    $ \Delta_{+} $ term in Eq. \ref{action} has scaling
   dimension 1/2, therefore is {\em relevant}, this term was first discovered
  by MF \cite{fisher1}.  The OPE of $ a_{s,i} $
   with itself will generate the dimension 2 energy momentum tensor
  of this Majorana fermion $ a_{s,i}(0,\tau) \frac{ \partial a_{s,i}(0,\tau) }{\partial \tau} $;
  the OPE of this energy momentum tensor with $ a_{s,i} $ will generate a first order descendant
  field of this primary field with dimension 3/2 $ L_{-1} a_{s,i}(0,\tau) = 
   \frac{ \partial a_{s,i}(0,\tau) }{\partial \tau} $ \cite{des}. 
   $ \tilde{q} $ term is the leading irrelevant operator with scaling
   dimension 3/2, therefore contributes to
\begin{equation}
  C_{imp} \sim (\tilde{q})^{2} T\log T
\label{heat1}
\end{equation}
  Where $ \tilde{q} = \sqrt{\frac{1-\cos \theta}{2}} q
   + \sqrt{\frac{1+\cos \theta}{2}} y $.

   It is important to see there is a {\em marginal} operator
   $ \partial \Phi^{n}_{s}(0) $ in the { \em spin} space which changes the angle
   $ \theta $ in Eq.\ref{obi}. This operator is very {\em different} from
  the marginal operator $ \partial \Phi_{c}(0) $ in the charge sector
  which changes the angle $ \theta_{ph} $ in Eq.\ref{charge}. Combined with
  the leading irrelevant operator, it will always generate the dimension
   1/2 relevant operator. This indicates that the existence of the line of NFL fixed
   points and the existence of one relevant operator are intimately connected.

   However, from Eqs.\ref{final}, \ref{local1}, only $ q $ term  contributes
   to the impurity susceptibility
\begin{equation}
   \chi^{h}_{imp} \sim  q^{2} \log T
\label{sus1}
\end{equation}

  As shown in Ref.\cite{line}, the Wilson Ration $ R=8/3 $ is universal for any
 {\em general } spin anisotropic 2CK. However, near this  line of NFL fixed points , $ R $ 
  is {\em not} universal.

 $ \gamma_{1} $ term has dimension 2.
   $ \tilde{y} $ term has scaling dimension 5/2, in $ H^{\prime} $, it can be written as 
\begin{equation}
 \tilde{y}  :\widehat{a}(\tau) \frac{\partial \widehat{a}(\tau)}{\partial \tau}: a_{s,i}(0,\tau)
 \sim  :\tilde{b}_{sf,i}(0, \tau) \frac{\partial \tilde{b}_{sf,i}(0, \tau) }{\partial \tau}: a_{s,i}(0,\tau)
\label{later}
\end{equation}
  Where $ \tilde{y} = -\sqrt{\frac{1+\cos \theta}{2}} q
   + \sqrt{\frac{1-\cos \theta}{2}} y $.

   This operator content is totally consistent with the following CFT analysis
   on the special point $ \theta=\pi $:
   if $ \theta=\pi $, the fixed point symmetry is $ O(3) \times O(5) $,
   the  CFT analysis of Ref.\cite{line} can be borrowed.
   Under the restriction of $ Z_{2} $ and Time-Reversal symmetry \cite{atten},
   it is easy to see there is only one relevant
   operator $ \phi^{1} $ with scaling dimension 1/2 which can be identified
   as $ a_{s,i} $, two leading irrelevant
   operators with scaling dimension 3/2, one is the spin  0  operator
   $ T^{0}_{0} $ which is {\em Virasoro primary} and can be identified as $ q $ term,
   another is the spin 1 operator $ L_{-1} \phi^{1} = \frac{ d \phi^{1} }{ d \tau}  $
   which is the {\em first order Virasoro descendant}.

    By Eq.\ref{twist} and Eq.\ref{define}, the spin-0 and spin-1 leading irrelevant operators 
    in $ H^{\prime} $ become \cite{check} 
\begin{eqnarray}
  \tilde{q}  \widehat{a}(\tau) \widehat{b}(\tau)
    a_{s,i}(0, \tau) \tilde{b}_{s,i}(0,\tau) \sim 
   \tilde{b}_{sf,i}(0, \tau) a_{s,i}(0, \tau) \tilde{b}_{s,i}(0,\tau)   \nonumber   \\
 = b_{s,i}(0,\tau) b_{sf}(0, \tau) a_{s,i}(0, \tau)
 = b^{\prime}_{s}(0,\tau) b_{sf}(0, \tau) a^{\prime}_{s}(0, \tau)     \nonumber   \\
   \frac{\partial a_{s,i}(0,\tau) }{\partial \tau} \sim 
   \frac{\partial a^{\prime}_{s}(0,\tau) }{\partial \tau}  ~~~~~~~~~~~~~~~~~~~~~~
\end{eqnarray}
 
    By Eq.\ref{general}, the corresponding spin-0 and spin-1 operators in $ H $ are
\begin{equation}
  \tilde{q} b_{s}(0,\tau)( a_{s}(0, \tau) b_{sf}(0, \tau)),~~~
   \frac{\partial b_{s}(0,\tau) }{\partial \tau}  
\end{equation}

    Obviously, only the coefficient $\tilde{q} $ {\em depend} on the angle 
 $\theta $ which specifies the position on the  line of NFL fixed points .

    They can be written in terms of the {\em new} bosons introduced in Eq.\ref{jinwu},
\begin{equation}
    \cos \Phi^{n}_{sf}(0,\tau) \partial \Phi^{n}_{s}(0,\tau),~~~
   \frac{\partial }{\partial \tau} \cos \Phi^{n}_{sf}(0,\tau)
\label{leading}
\end{equation}

  It is easy to see that the marginal operator $ \partial \Phi^{n}_{s}(0) $ makes
  no contribution to the Green function.

  Following Ref.\cite{powerful}, the first order correction to the single particle L-R Green
  functions due to the spin-0 operator can be
  calculated ( $ x_{1} >0, x_{2} <0 $ )
\begin{eqnarray}
\langle \psi_{1 +}( x_{1},\tau_{1} ) \psi^{\dagger}_{1 +}( x_{2},\tau_{2} ) \rangle  \sim
 \tilde{q} \int d \tau \langle \psi_{1+}(x_{1}, \tau_{1})
    \cos \Phi^{n}_{sf}(0,\tau) \partial \Phi^{n}_{s}(0,\tau)
    \psi^{\dagger}_{1+}(x_{2}, \tau_{2}) \rangle    \nonumber  \\
 \sim  \tilde{q} \int d\tau
 \langle e^{-\frac{i}{2} \Phi_{c}( x_{1}, \tau_{1} )} e^{\frac{i}{2} \Phi_{c}( x_{2}, \tau_{2} )}\rangle   
 \langle e^{-\frac{i}{2} \Phi^{n}_{s}( x_{1}, \tau_{1} )} \partial \Phi^{n}_{s}(0, \tau)
 e^{\frac{i}{2} ( \Phi^{n}_{s}( x_{2}, \tau_{2} ) + \theta )}\rangle   \nonumber   \\
 \times   \langle  e^{-\frac{i}{2} \Phi_{f}( x_{1}, \tau_{1} )} e^{\frac{i}{2} \Phi_{f}( x_{2}, \tau_{2} )}\rangle
 \langle e^{-\frac{i}{2} \Phi^{n}_{sf}( x_{1}, \tau_{1} )} e^{ i \Phi^{n}_{sf}( 0, \tau )} 
 e^{-\frac{i}{2} \Phi^{n}_{sf}( x_{2}, \tau_{2} )}\rangle   \nonumber   \\
  \sim  \tilde{q} e^{i \theta/2}  ( z_{1}- \bar{z}_{2})^{-3/2} ~~~~~~~~~~~~~~~~~~~~~~~~~~~~~~~~~~~
\label{sin}
\end{eqnarray}
    where $ z_{1}=\tau_{1}+i x_{1} $ is in the upper half plane,
     $ \bar{z}_{2}=\tau_{2}+i x_{2} $ is in the lower half plane.

  The first order correction to the single particle L-R Green
  functions due to the spin-1 operator
   vanishes because  the spin-1 leading irrelevant operator can be written
  as a total derivative and the three point functions are the periodic function of
  the imaginary time.

   The bosonized form of the $\tilde{y} $ term in Eq.\ref{later} in $ H $ is \cite{bose}
\begin{eqnarray}
  \tilde{y} [ \cos \theta :\cos 2\Phi^{n}_{s}(0,\tau): -\frac{1}{2} :( \partial \Phi^{n}_{s}(0,\tau))^{2}: 
                             ~~~~~~~~~~\nonumber  \\
  -\sin \theta (:\sin 2\Phi^{n}_{s}(0,\tau): - \partial^{2} \Phi^{n}_{s}(0,\tau)) ]
  \cos \Phi^{n}_{sf}(0,\tau)
\end{eqnarray}

   The first order correction due to this dimension 5/2 operator is
\begin{eqnarray}
\langle \psi_{1 +}( x_{1},\tau_{1} ) \psi^{\dagger}_{1 +}( x_{2},\tau_{2} ) \rangle \sim
      ~~~~~~~~~~~~~~~~~~~~~~~~~~~~\nonumber  \\
  \tilde{y} \int d\tau   
\langle e^{-\frac{i}{2} \Phi^{n}_{s}( x_{1}, \tau_{1} )} [-\frac{1}{2} 
:(\partial \Phi^{n}_{s}(0, \tau))^{2}:+\sin \theta \partial^{2} \Phi^{n}_{s}(0,\tau) ]
 e^{\frac{i}{2} ( \Phi^{n}_{s}( x_{2}, \tau_{2} ) + \theta )}\rangle   \nonumber   \\
 \times \langle e^{-\frac{i}{2} \Phi^{n}_{sf}( x_{1}, \tau_{1} )} e^{ i \Phi^{n}_{sf}( 0, \tau )} 
 e^{-\frac{i}{2} \Phi^{n}_{sf}( x_{2}, \tau_{2} )}\rangle   
  \sim   \tilde{y} e^{i \theta/2} (i C_{1} + C_{2} \sin \theta )  ( z_{1}- \bar{z}_{2})^{-5/2} ~~~~~~
\label{cos}
\end{eqnarray}
    Where $ C_{1}, C_{2} $ are real numbers.

  From Eq.\ref{leading}, it is easy to identify another dimension 5/2 operator
\begin{equation}
  : ( \partial \Phi^{n}_{s}(0,\tau))^{2}: \cos \Phi^{n}_{sf}(0,\tau)
\label{onemore}
\end{equation}

   The contribution from this operator can be similarly calculated.

  By using the following OPE:
\begin{eqnarray}
 : e^{-\frac{i}{2} \Phi( z_{1} )}: : e^{\frac{i}{2}  \Phi( z_{2})}: =
 (z_{1}-z_{2})^{-1/4}-\frac{i}{2}(z_{1}-z_{2})^{3/4} :\partial \Phi(z_{2}):  
                            \nonumber   \\
 -\frac{i}{4}(z_{1}-z_{2})^{7/4} :\partial^{2} \Phi(z_{2}):
 -\frac{1}{8}(z_{1}-z_{2})^{7/4} : (\partial \Phi(z_{2}) )^{2}: + \cdots
\label{ope}
\end{eqnarray}

   We get the three point functions
\begin{eqnarray}
 \langle  e^{-\frac{i}{2} \Phi( z_{1} )} \partial \Phi(z)  e^{\frac{i}{2}  \Phi( z_{2})} \rangle
  & = & \frac{-i/2}{ (z_{1}-z)(z_{1}-z_{2})^{-3/4}(z-z_{2})}   \nonumber   \\
 \langle  e^{-\frac{i}{2} \Phi( z_{1} )} \partial^{2} \Phi(z)  e^{\frac{i}{2}  \Phi( z_{2})} \rangle
  & = & \partial_{z}[ \frac{-i/2}{ (z_{1}-z)(z_{1}-z_{2})^{-3/4}(z-z_{2})} ]
\label{odd}
\end{eqnarray}

   By Conformal Ward identity, we can write the three point function with the energy momentum
   tensor
\begin{eqnarray}
 \langle  e^{-\frac{i}{2} \Phi( z_{1} )} T(z)  e^{\frac{i}{2}  \Phi( z_{2})} \rangle
   = \frac{1/8}{ (z_{1}-z)^{2}(z_{1}-z_{2})^{-7/4}(z-z_{2})^{2}}
\label{even}
\end{eqnarray}

  In Ref.\cite{affleck2}, AL found that in the multi-channel Kondo model,
  the electron self-energy has both real and imaginary
  parts which are {\em non-analytic } function of the frequency $ \omega $.
  In the presence of P-H symmetry, the imaginary part is {\em even } function of $ \omega $,
  the real part is {\em odd} function of $ \omega $,
  because the two parts are related by Kramers-Kronig relation. Only the part
  of self-energy which is both {\em imaginary} and {\em even} function of $ \omega $ contributes
  to the electron conductivity. The factor $ i $ will interchange {\em real } and {\em imaginary}
  part. In evaluating Eq.\ref{cos}, we used the important fact that the two three point functions
  in Eqs.\ref{odd},\ref{even} differ by the factor $ i $.

  By conformal transforming Eq.\ref{sin} to finite temperature,
  we get the {\em leading} term of the low temperature conductivity from channel one
  and parity $ + $ fermions
\begin{equation}
   \sigma_{1 +}(T) \sim \frac{\sigma_{u}}{2} (1- \tilde{q} \sin \frac{\theta}{2} \sqrt{T}),~~~~
   \sigma_{u}= \frac{ 2 \pi (e \rho_{F} v_{F} )^{2} }{ 3 n_{i} }
\label{sign1}
\end{equation}
    Where $  \rho_{F} $ is the density of states {\em per spin per channel} at the fermi energy,
   $ n_{i} $ is the impurity density.

     Similarly \cite{ambi}, we get the {\em leading} term of the low temperature conductivity
   from channel one and parity $ - $ fermions
\begin{equation}
   \sigma_{1 -}(T) \sim \frac{\sigma_{u}}{2} (1- \tilde{q} \sin \frac{\theta}{2} \sqrt{T})
\label{sign2}
\end{equation}

   Because of the global $ SU(2)_{f} $ symmetry in the flavor sector, the same equations hold in
   channel 2.

    Even the symmetry in the spin sector is $ O(1) \times U(1) $ instead of $ O(3) $ of the
 2CK, the 2 channel and 2 parity fermions do make the {\em same} leading contribution to the total conductivity
\begin{equation}
   \sigma(T) \sim 2\sigma_{u} (1- \tilde{q} \sin \frac{\theta}{2} \sqrt{T})
\label{resist1}
\end{equation}

 For $ \theta= \pi $, namely at the 2CK, $\tilde{q}=q $, then two universal ratios can be formed
 from Eqs. \ref{heat1},\ref{sus1},\ref{resist1}.

 For $ \theta=0 $, namely at the 2IK, $ \tilde{q}=y $, the coefficient of $ \sqrt{T} $ vanishes.
 It is evident that the 2nd order correction ( actually any {\em even} order  correction )
 to the Green function vanishes, the 3rd order correction will lead to $ T^{3/2} $,
 but the coefficients still vanishes due to $\sim \sin \theta/2=0 $, because {\em odd} order
 corrections have the same $ i $ factor.  

  By conformal transforming Eq.\ref{cos} to finite temperature \cite{connect},
  we get the {\em next-leading} term of the low temperature electrical conductivity 
\begin{equation}
   \sigma_{1+} \sim  \tilde{y}( \cos \frac{\theta}{2} C_{1} + \sin \frac{\theta}{2} \sin \theta C_{2}) T^{3/2}
\label{higher}
\end{equation}
  
    Putting $ \theta =0 $ ( then $\tilde{y}=-q $ ) in the above equation
    and adding the contribution from the operator in Eq.\ref{onemore},
    we get the leading term at the 2IK fixed point:
\begin{equation}
   \sigma(T) \sim 2 \sigma_{u} (1+  T^{3/2})
\label{resist11}
\end{equation}

   It is evident that even at the 2IK point, no universal ratios can be formed.

 For general $\theta $, the leading low temperature behaviors of the three physical measurable
 quantities are given by Eqs.\ref{heat1},\ref{sus1},\ref{resist1}, no universal ratios can be
 formed.

   The potential scattering term $ V_{1} $ is a marginal operator  which
   causes a phase shift in the charge sector:
\begin{equation}
   \Phi_{c,L}= \Phi_{c,R} + \theta_{ph}
\label{charge}
\end{equation}

   The symmetry of the fixed point is reduced to $ O(1) \times U(1) \times O(3) \times U(1) $,
   Eqs.\ref{sign1}, \ref{sign2} become:
\begin{eqnarray}
   \sigma_{1 +}(T) \sim \frac{\sigma_{u}}{2} (1- \tilde{q} \sin \frac{\theta +\theta_{ph}}{2} \sqrt{T})    \nonumber  \\
   \sigma_{1 -}(T) \sim \frac{\sigma_{u}}{2} (1- \tilde{q} \sin \frac{\theta-\theta_{ph}}{2} \sqrt{T})
\end{eqnarray}
   
  It is easy to see that in the presence of P-H symmetry breaking, {\em in contrast to}
 the 2CK, the different parity fermions do make different contributions to the
 conductivity, Eq.\ref{resist1} becomes
\begin{equation}
   \sigma(T) \sim 2\sigma_{u} (1- \tilde{q} \sin \frac{\theta}{2} \cos \frac{\theta_{ph}}{2} \sqrt{T})
\label{times}
\end{equation}
    
    Eq.\ref{higher} becomes 
\begin{eqnarray}
   \sigma_{1+} & \sim &  \tilde{y}( \cos \frac{\theta + \theta_{ph}}{2} C_{1} +
    \sin \frac{\theta+\theta_{ph}}{2} \sin \theta C_{2}) T^{3/2}   \nonumber   \\
   \sigma_{1-} & \sim &  \tilde{y}( \cos \frac{\theta - \theta_{ph}}{2} C_{1} +
    \sin \frac{\theta-\theta_{ph}}{2} \sin \theta C_{2}) T^{3/2}   
\end{eqnarray}

     The total conductivity becomes
\begin{equation}
   \sim  \tilde{y} \cos \frac{\theta_{ph}}{2}( \cos \frac{\theta}{2} C_{1} + \sin \frac{\theta}{2} \sin \theta C_{2}) T^{3/2}
\end{equation}

     The total conductivity at the 2IK becomes
\begin{equation}
   \sigma(T) \sim 2 \sigma_{u} (1+ \cos \frac{\theta_{ph}}{2} T^{3/2})
\end{equation}

      As shown by AL \cite{affleck1}, P-H symmetry breaking does not change the leading results of the
 specific heat and susceptibility.
  
       In this section, we discussed two marginal operators, one in spin space $ \partial \Phi^{n}_{s}(0) $,
  another in the charge space $ \partial \Phi_{c}(0) $. Both make {\em no} contributions
   to the conductivity and make only subleading contributions to the thermodynamic
   quantities. However, $ \partial \Phi^{n}_{s}(0) $ is much more important.
   Combined with the leading irrelevant operator in the spin sector, it will
   generate the dimension 1/2 relevant operator which will always make the line
   of NFL fixed points unstable. Furthermore, as shown by Eqs.\ref{heat1},\ref{sus1},
   even the coefficients of the leading terms of the thermodynamic quantities
   depend on the position on the line caused by $ \partial \Phi^{n}_{s}(0) $.

\section{Additional NFL Fixed point }
 If $ \tilde{q}=\tilde{y} =\Delta_{K} =0 $,
  this fixed point is located
 at $ \gamma_{1}=1, \gamma_{2}=0 $ where $\widehat{a} $ decouples,
  $ \widehat{b} $ loses its kinetic energy and becomes a
  Grassmann Lagrangian multiplier, integrating $ \widehat{b} $ out leads to
  the {\em boundary conditions} :

\begin{equation}
 a^{L}_{s,i}(0)=-a^{R}_{s,i}(0)
\end{equation}

    We also have the trivial boundary conditions
\begin{equation}
  b^{L}_{s,i}(0)= b^{R}_{s,i}(0),
  ~~~~~  b^{L}_{sf}(0)= b^{R}_{sf}(0) 
\end{equation}

  Using Eq.\ref{general}, the above boundary conditions of $ H^{\prime} $
  correspond to the following boundary conditions of $ H $: 
\begin{equation}
 a^{s}_{L}(0)  = -a^{s}_{R}(0) ,~~~ b^{s}_{L}(0)  =  b^{s}_{R}(0) ,~~~
 b^{sf}_{L}(0) =  b^{sf}_{R}(0) 
\label{twofold}
\end{equation}

   The above boundary conditions can be expressed in terms of the {\em new}
 chiral bosons in Eq.\ref{jinwu}:
\begin{equation}
\Phi^{n}_{s,L}(0) = - \Phi^{n}_{s,R}(0) + \pi
\end{equation}

   In terms of {\em new} physical fermions, it reads:
\begin{equation}
\psi_{ i \pm,L}(0) = e^{ \pm i \frac{\pi}{2} }  S_{i \pm,R}(0)
\end{equation}

    This is a {\em NFL  fixed point } with $ g=\sqrt{2} $ and the
   symmetry $ O(1) \times O(7) $ which is the fixed point symmetry of the 2IK(a)
  (Fig.~\ref{picture}).  At this fixed point, the original electrons scatter into
  the collective excitations which fit into the $ S $ spinor representation of SO(8).
  The finite size spectrum is listed in Table \ref{2IK}.

  The local correlation functions at this fixed point are
\begin{equation}
\langle \widehat{b}(\tau) \widehat{b}(0) \rangle \sim \frac{1}{\tau}, ~~~~
\langle a_{s,i}(0, \tau) a_{s,i}(0,0) \rangle \sim \frac{\gamma^{2}_{2}}{\tau^{3}}
\label{local2}
\end{equation}

   We can also read the scaling dimensions of the various fields
  $ [\widehat{a}]=0, [\widehat{b}]=1/2, [b_{s,i}]=[b_{sf}]=1/2, [a_{s,i}]=3/2 $.

      This NFL fixed point is more unlikely to be observed by experiments, because
  it has {\em two} relevant directions $ \Delta_{1},\Delta_{-} $.

  Similar to the discussions in the last section, at the 2IK fixed point,
  $ \widehat{a} $ decouples and $ \widehat{b} $ turns into
  the {\em non-interacting} scaling field at the fixed point
\begin{equation}
      \widehat{b}(\tau) \sim  a_{s,i}(0, \tau)
\end{equation}

    The corresponding scaling field in $ H $ is
\begin{equation}
     -b_{s}(0,\tau)
\end{equation}

     The impurity spin turns into
\begin{eqnarray}
  S_{x}(\tau) & \sim & i \widehat{a} a_{s} + \cdots     \nonumber  \\  
  S_{y}(\tau) & \sim & i( \widehat{a} b_{s} +a_{s} b_{s} ) +\cdots    \nonumber  \\ 
  S_{z}(\tau) & \sim & i \widehat{a}b_{s} +\cdots
\end{eqnarray}

  The impurity spin-spin correlation function $ \langle S^{a}(\tau) S^{a}(0) \rangle
  \sim 1/\tau $.

   On the  line of NFL fixed points  discussed in the previous section, if $ \theta =0$,
   the fixed point symmetry is enlarged to the 2IK(b) ( Fig.~\ref{picture} ).
   Although these two fixed points
   have the same symmetry, therefore the same finite size spectrum, 
   but the {\em allowed} boundary operators are very {\em different}.
   This additional NFL fixed point is also {\em unstable}.
   $ \Delta_{1}, \Delta_{-} $ are {\em two} relevant terms with scaling
   dimension 1/2, any linear combination of the two terms will
   drive the system to a given point on the line of FL fixed points to be discussed
   in the following section. They will generate two dimension $ 3/2 $
   leading irrelevant operators $L_{-1} b_{sf}(0,\tau), L_{-1}b_{s,i}(0,\tau) $
   respectively \cite{des} and two dimension 2 subleading irrelevant operators
   (energy-momentum tensors )
   $ b_{sf}(0,\tau) \frac{ \partial b_{sf}(0,\tau) }{\partial \tau}
   , b_{s,i}(0,\tau) \frac{ \partial b_{s,i}(0,\tau) }{\partial \tau} $.
    As explained in the last section, because the two leading irrelevant operators
   are {\em first order Virasoro descendants}, they make {\em no} contributions.
   $\gamma_{2} $ term also has dimension 2. In all there are {\em three} dimension 2 subleading
    irrelevant operators which contribute $ C_{imp} \sim T $. In all, we have
\begin{equation}
  C_{imp} \sim T
\label{heat2}
\end{equation}
    
    From Eqs.\ref{final},\ref{local2}, we get the susceptibility
\begin{equation}
   \chi^{h}_{imp} \sim  q^{2} \log T
\label{sus2}
\end{equation}

    By Eq.\ref{define}, the leading and subleading irrelevant operators in $ H^{\prime} $ become 
\begin{eqnarray}
    \frac{\partial b_{sf}(0,\tau)}{\partial \tau},~~~
     \frac{\partial b^{\prime}_{s}(0,\tau)}{\partial \tau} ~~~~~~~~~~~~~    \nonumber   \\
    b_{sf}(0,\tau) \frac{ \partial b_{sf}(0,\tau) }{\partial \tau},~~~
    b^{\prime}_{s}(0,\tau) \frac{ \partial b^{\prime}_{s}(0,\tau) }{\partial \tau}   \nonumber  \\
  \gamma_{2}  \widehat{b}(\tau) \frac{\partial \widehat{b}(\tau)}{\partial \tau}
    = \gamma_{2}  a^{\prime}_{s}(0,\tau) 
   \frac{\partial a^{\prime}_{s}(0,\tau)}{\partial \tau}
\end{eqnarray}

    By Eq.\ref{general}, the corresponding operators in $ H $ are
\begin{eqnarray}
    \frac{\partial b_{sf}(0,\tau)}{\partial \tau},~~~ \frac{\partial a_{s}(0,\tau)}{\partial \tau} ~~~~~~~~~~~~~~~~~~~~~
            \nonumber   \\
    b_{sf}(0,\tau) \frac{ \partial b_{sf}(0,\tau) }{\partial \tau},~~~~
    a_{s}(0,\tau) \frac{ \partial a_{s}(0,\tau) }{\partial \tau},~~~
    \gamma_{2}  b_{s}(0, \tau) \frac{\partial b_{s}(0,\tau)}{\partial \tau}
\label{twofold1}
\end{eqnarray}

    They can be written in terms of the {\em new} bosons \cite{omit} 
\begin{eqnarray}
    \frac{\partial}{\partial \tau} \sin \Phi^{n}_{s}(0,\tau),~~~ 
    \frac{\partial}{\partial \tau} \cos \Phi^{n}_{s}(0,\tau)    \nonumber   \\
     \pm \cos 2\Phi^{n}_{s}(0,\tau)- \frac{1}{2} :( \partial \Phi^{n}_{s}(0,\tau))^{2}:  \nonumber  \\
    \gamma_{2} ( \cos 2\Phi^{n}_{sf}(0,\tau)-
    \frac{1}{2} :( \partial \Phi^{n}_{sf}(0,\tau))^{2}:)
\label{change}
\end{eqnarray}

    The first order corrections to the single particle L-R Green function
    due to the 1st and the 2nd operators in Eq.\ref{change} vanish. 

  Those due to the 3rd and 4th operators are \cite{omit}
\begin{eqnarray}
 \langle  \psi_{1 +}( x_{1},\tau_{1} ) \psi^{\dagger}_{1 +}( x_{2},\tau_{2} ) \rangle
        \sim  ~~~~~~~~~~~~~~~~~~~~~
                                  \nonumber  \\
 \int d \tau   \langle e^{-\frac{i}{2} \Phi^{n}_{s}( x_{1}, \tau_{1} )}
     ( \pm \cos 2\Phi^{n}_{s}(0,\tau) -\frac{1}{2} :( \partial \Phi^{n}_{s}(0,\tau))^{2}:)
 e^{\frac{i}{2} (- \Phi^{n}_{s}( x_{2}, \tau_{2} )+ \pi)}\rangle =0  
\end{eqnarray} 

     The 5th operator in Eq.\ref{change} makes {\em no} contributions to the Green function either:
\begin{eqnarray}
& \langle & \psi_{1 +}( x_{1},\tau_{1} ) \psi^{\dagger}_{1 +}( x_{2},\tau_{2} ) \rangle  \sim
 \gamma_{2} \int d \tau
 \langle e^{\frac{i}{2} (-\Phi^{n}_{s}( x_{1}, \tau_{1} ) +\pi )}
 e^{-\frac{i}{2}  \Phi^{n}_{s}( x_{2}, \tau_{2} ) }\rangle   \nonumber   \\
 & \times &   \langle e^{-\frac{i}{2} \Phi^{n}_{sf}( x_{1}, \tau_{1} )}
     ( \cos 2\Phi^{n}_{sf}(0,\tau) -\frac{1}{2} :( \partial \Phi^{n}_{sf}(0,\tau))^{2}:)
 e^{\frac{i}{2}  \Phi_{sf} ( x_{2}, \tau_{2} )}\rangle =0   
\end{eqnarray}

  It is easy to see higher order corrections due to the above operators also vanish.

   The  $ y $ and $ q $ terms are two irrelevant operators
    with scaling dimension 5/2, they can be written in $ H^{\prime} $ as
\begin{equation}
   : \widehat{b}(\tau) \frac{\partial \widehat{b}(\tau)}{\partial \tau}: b_{sf}(0,\tau),~~~
   : \widehat{b}(\tau) \frac{\partial \widehat{b}(\tau)}{\partial \tau}: b_{s,i}(0,\tau)
\end{equation}

   The bosonized forms in $ H $ are
\begin{eqnarray}
   ( \cos 2\Phi^{n}_{sf}(0,\tau)- \frac{1}{2} :( \partial \Phi^{n}_{sf}(0,\tau))^{2}:)
     \sin \Phi^{n}_{s}(0,\tau)    \nonumber  \\
   ( \cos 2\Phi^{n}_{sf}(0,\tau)- \frac{1}{2} :( \partial \Phi^{n}_{sf}(0,\tau))^{2}:)
     \cos \Phi^{n}_{s}(0,\tau)  
\end{eqnarray}

   The first order correction due to the first operator is \cite{connect}
\begin{eqnarray}
& \langle & \psi_{1 +}( x_{1},\tau_{1} ) \psi^{\dagger}_{1 +}( x_{2},\tau_{2} ) \rangle  \sim
 i \int d \tau
 \langle e^{-\frac{i}{2} \Phi^{n}_{s}( x_{1}, \tau_{1} )} e^{i \Phi^{n}_{s}(0,\tau)}
 e^{\frac{i}{2} (- \Phi^{n}_{s}( x_{2}, \tau_{2} ) +\pi) }\rangle   \nonumber   \\
 & \times &   \langle e^{-\frac{i}{2} \Phi^{n}_{sf}( x_{1}, \tau_{1} )}
      :( \partial \Phi^{n}_{sf}(0,\tau))^{2}:
 e^{\frac{i}{2}  \Phi_{sf}( x_{2}, \tau_{2} )}\rangle 
  \sim i (z_{1}-\bar{z}_{2})^{-5/2}
\label{save}
\end{eqnarray}
   The above integral is essentially the same as the first part of that in Eq.\ref{cos}.

  As explained in the preceding section, due to the factor $ i $ difference than the first
  operator, the 2nd operator makes {\em no} contribution to the conductivity .

   The other two dimension 5/2 operators are \cite{flavor}
\begin{equation}
  L_{-2} b_{sf}(0,\tau) = \frac{3}{4} \frac{\partial^{2}}{\partial \tau^{2}} \sin \Phi^{n}_{s}(0,\tau),~~~ 
  L_{-2} a_{s}(0,\tau) = \frac{3}{4}  \frac{\partial^{2}}{\partial \tau^{2}} \cos \Phi^{n}_{s}(0,\tau)   
\end{equation}
  
     Because they can still be written as total derivatives, therefore make no contributions to
  the Green functions.

   Conformal transformation of Eq.\ref{save} to finite temperature and scaling analysis \cite{twoimp}
 lead to
\begin{equation}
  \sigma(T) \sim 2\sigma_{u} (1+  T^{3/2})
\label{resist2}
\end{equation}

   There is {\em no} chance to form universal ratios from Eqs.\ref{heat2},\ref{sus2},\ref{resist2}.

  In Appendix C, similar calculations in the {\em old} boson basis Eq.\ref{second} are performed.

\section{ The line of FL fixed points }
 If $ \tilde{q}=\tilde{y} =0 $,
  this fixed point is located
 at $ \gamma_{1}= \gamma_{2}=0 $ where both $\widehat{a} $ and
   $ \widehat{b} $ lose their kinetic energies and become two
  Grassmann Lagrangian multipliers, integrating them out leads to
  the following {\em boundary conditions} :

\begin{equation}
 a^{L}_{s,i}(0)=-a^{R}_{s,i}(0),
  ~~~~~ \tilde{b}^{L}_{sf,i}(0)= -\tilde{b}^{R}_{sf,i}(0) 
\end{equation}

    We also have the trivial boundary conditions
\begin{equation}
 \tilde{b}^{L}_{s,i}(0)= \tilde{b}^{R}_{s,i}(0) 
\end{equation}

 Following the same procedures as those in the discussion of
  the  line of NFL fixed points , the above boundary
 conditions correspond to the boundary conditions in $ H $
\begin{equation}
 b^{s}_{L}(0)  =  b^{s}_{R}(0), ~~~~~ \psi^{a}_{L}(0) =  e^{i \theta} \psi^{a}_{R}(0) 
\label{vectorf}
\end{equation}

  The above boundary condition can be expressed in terms of
  the {\em new} bosons in Eq.\ref{jinwu}
\begin{equation}
 \Phi^{n}_{s,L}=\Phi^{n}_{s,R}+ \theta
\label{fermi}
\end{equation}

    As discussed in the  line of NFL fixed points , the physical fermions transform as
   the spinor representation
  of the above boundary conditions, the corresponding boundary conditions are
  derived in Appendix B, the result is
\begin{equation}
 \psi^{L}_{i \pm }  =   e^{  \pm i \theta/2 } \psi^{R}_{i \pm}
\label{last}
\end{equation}
 
   It can be checked explicitly the above boundary conditions
  satisfy all the symmetry requirement.
  The fermion fields with the even and the odd parity under $ Z_{2} $
   suffer opposite continuously changing phase shifts
   $ \pm \frac{\theta}{2} $
   along this line of fixed points . Depending on the sign of $ \Delta_{0} $, the impurity
   will occupy either the  even parity state with $ S^{x} =\frac{1}{2} $ or
   the odd parity state $ S^{x}= -\frac{1}{2} $. This simple physical picture
   should be expected from the starting Hamiltonian Eq.\ref{start}:
   If we keep $ \Delta_{0} $ term fixed, then $ V_{3} $ term is irrelevant,
   the $ y $ term ( $ y=2 V_{2} $ ) is exact marginal, it causes
   continuous opposite phase shifts for $ \psi_{i \pm} $.
   $ V_{3} $ term will generate
   the dimension 2 operator $ : J^{z}(0)J^{z}(0) : $, the OPE of the $ y $ term
   and the $ V_{3} $ term will generate the $ S^{z}J^{y}(0) $ term, this term will
   generate {\em another} dimension 2 operator $ : J^{y}(0) J^{y}(0) : $. 
   The impurity spin-spin correlation function $ S^{z}(\tau) S^{z}(0)
   \sim ( \frac{ V_{3}}{ \Delta_{0}})^{2} \frac{1}{ \tau^{2}} $ which is
    consistent with Eq.\ref{kick}.

   One particle matrix is $ S^{\pm}_{1} = e^{ \pm i \theta/2 } $,
   the residual conductivity $ \sigma(0) = 2\sigma_{u}/(1- \cos \frac{\theta}{2} ) $.
   This is a {\em  line of FL fixed points  } with $ g=1 $ and the symmetry $ U(1)
   \times O(6) \sim U(4) $ which interpolates continuously between the non-interacting
   fixed point and the 2CSFK fixed point found by the author
   in Ref. \cite{spinflavor}.
   If $ \theta=0 $, namely $ \Delta_{1}=0 $, the residual conductivity comes from
   the potential scattering term discussed in Sec. V which is omitted in this section,
   the fixed point
   symmetry is enlarged to $ O(8) $ which is the fixed point symmetry of the
   non-interacting electrons. We can see this very intuitively from Eq. \ref{start}:
   if $ y= \Delta_{1} = \Delta_{2}=0 $, then they will remain zero, the impurity will
   be in either the even parity state or the odd parity state depending on the sign
    of $ \Delta_{0} $, the electrons and the impurity
   will asymptotically decouple, therefore the electrons are asymptotically free \cite{deg}.
   Non-zero $ \Delta_{2} $ will {\em not} change the above physical picture, 
   because it will not generate $ y $ and $ \Delta_{1} $ terms, however
   if $ \Delta_{+} = \Delta_{0} + \Delta_{2}=0 $, the 2IK(b) is reached,
   if $ \Delta_{-} = \Delta_{0} - \Delta_{2}=0 $, the 2IK(a) is reached,
   As shown in Fig.\ref{picture}, both fixed points will flow to the free fixed point
   asymptotically. If $ \theta=\pi $, namely $ \Delta_{-}=0 $, $ \sigma(0)=2 \sigma_{u} $,
   the fixed point
   symmetry is enlarged to $ O(2) \times O(6) $ which is the the fixed point
   symmetry of the 2CSFK.  

   This line of FL fixed points is {\em stable}. There is {\em no} relevant operator.
    There is one marginal operator in the spin sector $ \partial \Phi^{n}_{s}(0) $
    which changes the angle $ \theta $ in Eq. \ref{fermi}. 
    $ \gamma_{1} $ and $\gamma_{2} $ terms which are leading irrelevant operators
    have dimension 2, they lead to typical fermi liquid behaviors.
    $ \tilde{q} $ term has dimension 3; in $ H^{\prime} $, it can be written as 
    $ : \widehat{b}(\tau) \frac{\partial \widehat{b}(\tau)}{\partial \tau} : 
    b_{s,i}(0,\tau) b_{sf}(0, \tau) $.
   The $ \tilde{y} $ term has dimension 4; in $ H^{\prime} $, it can be written as
   $ :\widehat{a}(\tau) \frac{\partial \widehat{a}(\tau)}{\partial \tau}:
   : \widehat{b}(\tau) \frac{\partial \widehat{b}(\tau)}{\partial \tau} : $.
   This operator contents
   are completely consistent with our direct analysis mentioned above.
    This complete agreement provides a very powerful check on the method developed
    in Ref.\cite{powerful}.

  The local correlation functions at this fixed point are
\begin{eqnarray}
\langle \widehat{a}(\tau) \widehat{a}(0) \rangle & \sim & \frac{1}{\tau}, ~~~~
\langle \tilde{b}_{sf,i}(0, \tau) \tilde{b}_{sf,i}(0,0) \rangle \sim \frac{\gamma^{2}_{1}}{\tau^{3}}
                                       \nonumber \\
\langle \widehat{b}(\tau) \widehat{b}(0) \rangle & \sim & \frac{1}{\tau}, ~~~~
\langle a_{s,i}(0, \tau) a_{s,i}(0,0) \rangle \sim \frac{\gamma^{2}_{2}}{\tau^{3}}
\end{eqnarray}

  From the above equation, we can also read the scaling dimension of the various fields
  $ [\widehat{a}]=[\widehat{b}]=1/2, [\tilde{b}_{s,i}]=1/2,
   [a_{s,i}]=[\tilde{b}_{sf,i}]=3/2 $.

   At the fixed point, $ \widehat{a}, \widehat{b} $ turn into
   the {\em non-interacting} scaling fields in $ H^{\prime} $
\begin{eqnarray}
      \widehat{a}(\tau) & \sim &  \tilde{b}_{sf,i}(0, \tau)=
       \frac{1}{\Delta_{K}}( -\Delta_{-} b_{s,i}(0,\tau)
          + \Delta_{1} b_{sf}(0,\tau))   \nonumber   \\ 
      \widehat{b}(\tau) & \sim &  a_{s,i}(0, \tau)
\end{eqnarray}

    The corresponding two scaling fields in $ H $ are
\begin{eqnarray}
      \frac{1}{\Delta_{K}}( -\Delta_{-} a_{s} + \Delta_{1} b_{sf})   \nonumber   \\ 
      -b_{s}(0, \tau)  ~~~~~~~~~~~~~~~
\end{eqnarray}

     The impurity spin turns into
\begin{eqnarray}
  S_{x}(\tau) &\sim &- i  \frac{\Delta_{1}}{\Delta_{K}}  a_{s} b_{sf} +\cdots  \nonumber  \\
  S_{y}(\tau) & \sim & i( a_{s} b_{s} + 
   \frac{1}{\Delta_{K}}( -\Delta_{-} a_{s} + \Delta_{1} b_{sf})b_{s} ) +\cdots   \nonumber  \\
  S_{z}(\tau) & \sim & i \frac{1}{\Delta_{K}}( -\Delta_{-} a_{s} + \Delta_{1} b_{sf})b_{s}  +\cdots 
\end{eqnarray}

  The impurity spin-spin correlation function shows typical FL behavior
\begin{equation}
  \langle S^{a}(\tau) S^{a}(0) \rangle \sim 1/\tau^{2}
\label{kick}
\end{equation}

     The two leading irrelevant operator in $ H $ become 
\begin{eqnarray}
  \gamma_{1}  \widehat{a}(\tau) \frac{\partial 
  \widehat{a}(\tau)}{\partial \tau} = ~~~~~~~~~~~~~~~~~~~~~~~~~
                              \nonumber    \\
    (\frac{\Delta_{-}}{\Delta_{K}})^{2} a_{s} \frac{\partial a_{s}(\tau)}{\partial \tau}
- 2\frac{ \Delta_{-} \Delta_{1}}{ \Delta^{2}_{K}} a_{s} \frac{\partial b_{sf}(\tau)}{\partial \tau}
 + (\frac{\Delta_{1}}{\Delta_{K}})^{2} b_{sf} \frac{\partial b_{sf}(\tau)}{\partial \tau} 
                     \nonumber  \\
  =  \cos \theta :\cos 2\Phi^{n}_{s}(0,\tau): -\frac{1}{2} :( \partial \Phi^{n}_{s}(0,\tau))^{2}: 
                             ~~~~~~~~~~\nonumber  \\
  -\sin \theta (:\sin 2\Phi^{n}_{s}(0,\tau): - \partial^{2} \Phi^{n}_{s}(0,\tau))~~~~~~~~~~~~~~~~~
                           \nonumber  \\
  \gamma_{2}  \widehat{b}(\tau) \frac{\partial \widehat{b}(\tau)}{\partial \tau}
   \sim \gamma_{2}( :\cos 2\Phi^{n}_{sf}(0,\tau): -\frac{1}{2} :( \partial \Phi^{n}_{sf}(0,\tau))^{2}:)
\label{win}
\end{eqnarray}
   
     Although the first operator do depend on the angle $\theta $, its scaling dimension
    remains 2, therefore will not affect the exponents of
   any physical measurable quantities.
  We refer the readers to Ref.\cite{flavor} for the detailed similar calculations on
  the single particle Green function and the electron conductivity .
    Second order corrections to the single particle Green functions from the two leading
  irrelevant operators lead to $ \sigma(T) \sim \sigma(0) + C(\theta) T^{2} $.

\section{ The effects of external magnetic field}

 According to Ref.\cite{fisher1}, the parameters in Eq.\ref{start} are
\begin{eqnarray}
   V_{1} &= & \pi \rho_{F} V  \\
   V_{2} &= & \pi \rho_{F} V \frac{\sin k_{F} R}{ k_{F} R}   \\
   V_{3} &= & \pi \rho_{F} V \sqrt{ 1- (\frac{\sin k_{F} R}{ k_{F} R})^{2} }
\end{eqnarray}  

   The external magnetic field $ H $ breaks the $ SU(2) $ flavor ( the real spin ) symmetry.
 it causes the energy band of spin $ \uparrow $ electrons to shift downwards, that of spin $\downarrow $
 to shift upwards. Channel 1 and 2 electrons have {\em different} fermi momenta, therefore
 couple to the impurity with {\em different} strength. Setting the external strain $ h=0 $,
 the Hamiltonian is

 \begin{eqnarray}
 H &= & H_{0} +i \delta v_{F} \int dx ( \psi^{\dagger}_{1 \alpha}(x)
  \frac{ \partial \psi_{1 \alpha}(x)}{ \partial x}
 - \psi^{\dagger}_{2 \alpha}(x) \frac{ \partial \psi_{2 \alpha}(x)}{ \partial x} )   \nonumber   \\
  &+ & V_{1} J_{c} (0) + + \delta V_{1} J^{z}_{f} (0) +
 2 V_{2} J^{x}(0) + 2 \delta V_{2} \tilde{J}^{x}(0)    
      \nonumber   \\
  & + & 4 V_{3} S^{z} J^{z}(0) +\Delta_{1}  (J^{x}(0) S^{x} + J^{y}(0) S^{y} )     \nonumber  \\
  &+ &  4 \delta V_{3} S^{z} \tilde{J}^{z}(0) + \delta \Delta_{1}  ( \tilde{J}^{x}(0) S^{x} + \tilde{J}^{y}(0) S^{y} )
       \nonumber   \\
  & + & \frac{ \Delta_{0} }{ \pi \tau_{c} } S^{x}
      +\Delta_{2} 2 \pi \tau_{c} ( S^{-}
  \psi^{\dagger}_{1 \uparrow} \psi_{1 \downarrow}
   \psi^{\dagger}_{2 \uparrow} \psi_{2 \downarrow}  +h.c.)
\label{field}
 \end{eqnarray}
    Where $ \tilde{J}^{a}(x)=J^{a}_{1}(x)-J^{a}_{2}(x) $ and all the $\delta $ terms are
      $ \sim H $.

 The term $ \frac{\delta v_{F}}{ 2 \pi} \int dx(
        \partial \Phi_{c}(x) \partial \Phi_{f}(x) +
      \partial \Phi_{s}(x) \partial \Phi_{sf}(x) ) $ does not couple to the impurity,
   therefore can be neglected.

    It is important to observe the bare hopping term and the two electron assisted hopping
  term are {\em not} affected by the magnetic field. Following Ref.\cite{flavor},
  the transformed Hamiltonian $ H^{\prime}= U H U^{-1} $ is
 \begin{eqnarray}
 H^{\prime}  &= & H_{0}+ 2 y \widehat{a} \widehat{b} a_{s,i}(0) b_{sf}(0)
  + 2 \delta y \widehat{a} \widehat{b} a_{sf}(0) b_{s,i}(0)   \nonumber   \\
   &+ & 2 q  \widehat{a} \widehat{b} a_{s,i}(0) b_{s,i}(0)
  + 2 \delta q  \widehat{a} \widehat{b} a_{sf}(0) b_{sf}(0)
                       \nonumber  \\
  & - & i \frac{ \Delta_{1}}{\sqrt{ 2 \pi \tau_{c} }} \widehat{a} b_{sf}(0)
   +  i \frac{ \delta \Delta_{1}}{\sqrt{ 2 \pi \tau_{c} }} \widehat{b} a_{sf}(0)
                        \nonumber  \\
  & +  & i \frac{ \Delta_{+}}{\sqrt{ 2 \pi \tau_{c} }} \widehat{b} a_{s,i}(0)
   +i \frac{ \Delta_{-}}{\sqrt{ 2 \pi \tau_{c} }} \widehat{a} b_{s,i}(0)
 \end{eqnarray}

    Performing the rotation Eq.\ref{twist}, the above equation can be rewritten as

\begin{eqnarray}
 H^{\prime} &  = &    H_{0}
  +  2 \tilde{q} \widehat{a} \widehat{b} a_{s,i}(0) \tilde{b}_{s,i}(0)
  +  2 \delta \tilde{q} \widehat{a} \widehat{b} a_{sf}(0) \tilde{b}_{sf,i}(0)  \nonumber  \\
  & + & 2 \tilde{y} \widehat{a} \widehat{b} a_{s,i}(0) \tilde{b}_{sf,i}(0)
  + 2 \delta\tilde{y} \widehat{a} \widehat{b} a_{sf}(0) \tilde{b}_{s,i}(0)
 \nonumber  \\
   & - &    i \frac{ \Delta_{K}}{\sqrt{ 2 \pi \tau_{c} }}
      \widehat{a} \tilde{b}_{sf,i}(0)
   +i \frac{ \Delta_{+}}{\sqrt{ 2 \pi \tau_{c} }} \widehat{b} a_{s,i}(0)
   +i \frac{ \delta \Delta_{1}}{\sqrt{ 2 \pi \tau_{c} }} \widehat{b} a_{sf}(0)
\label{sy}
\end{eqnarray}
   
   It is evident that the magnetic field $ H $ introduces, another relevant operator
 with scaling dimension 1/2. Under the combination of the two relevant directions in
 the above equation, the system flows to the line of FL fixed points with the boundary
 conditions
\begin{equation}
 \Phi^{n}_{s,L}=\Phi^{n}_{s,R}+ \theta_{s},~~~~ \Phi^{n}_{sf,L}=\Phi^{n}_{sf,R}+ \theta_{sf}
\end{equation}

    If $ H=0 $, $ \theta_{sf}=0 $, the boundary condition Eq.\ref{fermi} is recovered. 
    If $ \Delta_{+}=0 $, $ \theta_{sf}=\pi $. 

    There are two marginal operators along this line of FL fixed points,
    one $ \partial \Phi^{n}_{s}(0) $ is in the spin sector which
   changes the angle $ \theta_{s} $
   another $ \partial \Phi^{n}_{sf}(0) $ is in the spin-flavor  sector which
   changes the angle $ \theta_{sf} $. The corresponding boundary conditions
   in the original fermion basis can be similarly worked out as in the last section.

\section{ Scaling analysis of the physical measurable quantities }

   In this section, following the methods developed in Ref.\cite{qc,detail}
  and also considering the correction due to the {\em leading} irrelevant operators,
  we derive the scaling functions of the conductivity, impurity specific heat and 
  susceptibility :
\begin{equation}
   A(T, \Delta_{+}, H, \lambda) =  F( \frac{a \Delta_{+}}{\sqrt{T}}, \frac{b H}{\sqrt{T}}, \lambda \sqrt{T} )
\label{cond}
\end{equation}
    Where $ a,b $ are  non-universal metric factors which depend on $ \theta, \theta_{ph} $
  and the cutoff of the low energy theory, the dependence on $ \theta $ is due to the existence
  of the exactly marginal operator $ \partial \Phi^{n}_{s}(0) $ in the spin sector \cite{irre},
  the Kondo temperature is given by $ T_{K} \sim \lambda^{-2} $.

    We confine $ T < T_{K} $, so $ \lambda \sqrt{T} $ is a small parameter,
    we expand the right hand side of Eq.\ref{cond} in terms of the leading irrelevant operator
\begin{equation}
   A(T, \Delta_{+}, H, \lambda) = f_{0}( \frac{a \Delta_{+}}{\sqrt{T}}, \frac{b H}{ \sqrt{T}} ) +
        \lambda \sqrt{T} f_{1}( \frac{ a \Delta_{+}}{\sqrt{T}}, \frac{b H}{ \sqrt{T}} ) +
        (\lambda \sqrt{T})^{2} f_{2}( \frac{ a \Delta_{+}}{\sqrt{T}}, \frac{b H}{ \sqrt{T}} ) + \cdots
\label{expand}
\end{equation}

       For simplicity, we only consider $ \Delta_{+} \neq 0 $ or $ H \neq 0 $. The general
   case Eq.\ref{cond} can be discussed along the similar line of Ref.\cite{read}.

    From Eq.\ref{sy}, it is easy to observe that
    $ \Delta_{+} $ term and the magnetic field $ H $ term play very similar roles. In
    the following, we only explicitly derive the scaling function in terms of
    $ \Delta_{+} $. The scaling functions in the presence of $ H $ can be
    obtained by replacing $ \Delta_{+} $ by $ H $.

     As discussed in Sec. V, depending on the sign of $ \Delta_{+} $, the impurity is either in even parity
  or odd parity states,  but the physical measurable quantities
  should not depend on if the system flows to FL1 (even parity)
  or FL2 (odd parity ), so the above scaling function should only depend on $ |\Delta_{+}| $. 

   In the following, we discuss the scaling functions of $ \sigma(T)-\sigma(0), C_{imp}, \chi^{h}_{imp}
   ,\chi_{imp} $ respectively. Here $  \chi^{h}_{imp}, \chi_{imp} $ are the impurity
   hopping and spin susceptibility respectively.

{\em The scaling function of the conductivity} 
 
   We look at the two different limits of the $ f $ functions.

   Keeping $ T < T_{K} $ fixed, let $ \Delta_{+} \rightarrow 0 $, the system is in the Quantum Critical(QC)
   regime controlled by the line of NFL fixed points. We can do perturbative expansions in terms of 
  $ \Delta_{+}/\sqrt{T} $. As discussed in Refs.\cite{powerful,flavor},
   the {\em overall sign ambiguity} in the spinor representation
   of the NFL boundary condition Eq.\ref{obi} should be fixed by the requirement of
   symmetry and analyticity \cite{ambi}.
   The perturbative expansions are 
\begin{eqnarray}
  \sigma_{i +}(T, \Delta_{+}, \lambda=0 ) - \sigma(0) & \sim &  
   \frac{\Delta_{+}}{\sqrt{T}} + ( \frac{\Delta_{+}}{\sqrt{T}} )^{3} + \cdots    \nonumber  \\
  \sigma_{i -}(T, \Delta_{+}, \lambda=0 ) - \sigma(0)  & \sim &  
   -\frac{\Delta_{+}}{\sqrt{T}} - ( \frac{\Delta_{+}}{\sqrt{T}} )^{3} + \cdots
\label{cancel}
\end{eqnarray}

     The total conductivity $ \sigma(T, \Delta_{+}, \lambda=0 ) -\sigma(0) = 0 $, therefore
    $ f_{0}(x) \equiv 0 $. The conductivities
  from the different parities have to cancel each other, otherwise, we get a {\em non-analytic}
  dependence at {\em small} magnetic field at {\em finite} temperature \cite{thank}.
   
\begin{eqnarray}
    f_{1}( x ) & = & 1 + x^{2} +x^{4} + \cdots, ~~~~ x \ll 1
\end{eqnarray}
 
     Substituting the above equation into Eq.\ref{expand}, we get
 
\begin{equation}
   \sigma(T, \Delta_{+}, \lambda)-\sigma(0)  = 
    \lambda \sqrt{T} + \frac{ \lambda \Delta_{+}^{2}}{ \sqrt{T}} + \cdots   
\end{equation}

  Keep $ \Delta_{+} $ fixed, but small, let $ T \rightarrow 0 $, the system is in FL1 (or FL2) regime,
  the conductivity should reduce to the FL form
\begin{eqnarray}
    f_{0}( x ) & \equiv &  0   \nonumber  \\
    f_{1}( x ) & =  & |x| + |x|^{-3} + \cdots, ~~~~ x \gg 1    \nonumber   \\
    f_{2}( x ) & \equiv  & 0
\end{eqnarray}

       Substituting the above equation into Eq.\ref{expand}, we have 
\begin{equation}
   \sigma(T, \Delta_{+}, \lambda)-\sigma(0)  =   \lambda |\Delta_{+} | + (\lambda |\Delta_{+} | )^{3} +
                  +   ( \lambda |\Delta_{+}|^{-3} + \cdots ) T^{2}+\cdots
\label{test1}
\end{equation}

      The above equation indicates that the coefficient of $ T^{2} $ diverges
  as $ \lambda | \Delta_{+} |^{-3} $ instead of  $ \Delta_{+}^{-4} $ as we approach
  to the line of NFL fixed points.
  This means the relevant operator $ \Delta_{+} $ with scaling dimension 1/2 combined
  with the leading irrelevant operator $ \lambda $ with scaling dimension $ -1/2 $
  near the line of NFL fixed points will turn into one of the irrelevant operators $ \lambda_{FL,-2} $
  with scaling dimension $ -2 $ near the line of FL fixed points
\begin{equation}
     \lambda_{FL,-2} \sim \lambda |\Delta_{+}|^{-3}
\end{equation}
       
      First order perturbation in this operator leads to Eq.\ref{test1}.

{\em The scaling function of the impurity specific heat} 
 
     In the QC regime, the perturbative expansions give (up to possible logarithm):
\begin{eqnarray}
    g_{0}( x ) & = & x^{2} + x^{4} + \cdots, ~~~~ x \ll 1        \nonumber   \\
    g_{1}( x ) & \equiv &  0                                    \nonumber  \\
    g_{2}( x ) & = & 1 + x^{2} + \cdots, ~~~~ x \ll 1
\end{eqnarray}
      
      Substituting the above equation into Eq.\ref{expand}, we get
\begin{equation}
   C_{imp}  =  \frac{\Delta_{+}^{2}}{T} + \frac{\Delta_{+}^{4}}{T^{2}} + \lambda^{2} T \log T
    + \lambda^{2} \Delta_{+}^{2} + \cdots 
\end{equation}

     It was known that there are {\em accidental logarithmic violations} of scaling when the number of
    channel is two \cite{affleck1}. This violation has nothing to do with the existence of
    marginally irrelevant operators \cite{irre}. Similar violation occur in itinerant magnetism \cite{millis}.

  In the FL regime, the impurity specific heat should reduce to the FL form
\begin{eqnarray}
    g_{0}( x ) & = & x^{-2} + \cdots, ~~~~ x \gg 1    \nonumber   \\
    g_{2}( x ) & = &   c  + \cdots, ~~~~ x \gg 1
\end{eqnarray}
     
      Substituting the above equation into Eq.\ref{expand}, we get
\begin{equation}
   C_{imp}  =   T( \Delta_{+}^{-2} + \lambda^{2} + \cdots) + \cdots 
\label{test2}
\end{equation}

      The above equation indicates that the coefficient of $ T $ diverges
  as $ \Delta_{+}^{-2} $ as we approach to the line of NFL fixed points.
  This means the relevant operator $ \Delta_{+} $ with scaling dimension 1/2
  near the line of NFL fixed points will turn into one of the leading irrelevant operators $ \lambda_{FL,-1} $
  with scaling dimension $ -1 $ near the line of FL fixed points
\begin{equation}
     \lambda_{FL,-1} \sim \Delta_{+}^{-2}
\end{equation}

      First order perturbation in this operator leads to Eq.\ref{test2}. However, as shown
  in Eq.\ref{test1}, this leading irrelevant operator make {\em no} contribution  to the {\em total}
  conductivity, even though it makes contribution to even and odd parity conductivity separately
  (see Eq.\ref{cancel} ).

{\em The scaling function of the impurity hopping susceptibility } 
 
    In the QC regime, the perturbative expansions give
    (up to possible logarithm)
\begin{equation}
   \chi^{h}_{imp}  =  \lambda^{2} \log T + \frac{ \lambda^{2} \Delta_{+}^{2} }{T} + \cdots
\end{equation}

    In the FL regime   
\begin{equation}
   \chi^{h}_{imp}  =  \lambda^{2} \log 1/\Delta_{+}^{2} + \cdots
\end{equation}

     The exact crossover function can be calculated along the EK line in Eq.\ref{final}.
     In the FL regime, the Wilson Ratio $ R= T \chi^{h}_{imp}/ C_{imp} \sim \lambda^{2}
        \Delta_{+}^{2} \log 1/ \Delta^{2}_{+} $ is very small as $ \Delta_{+} \rightarrow 0 $. 

{\em The scaling function of the impurity spin susceptibility} 
   
    In this part, we set $ \Delta_{+}=0 $ and consider finite $ H $.

    In the QC regime, the perturbative expansions give (up to possible logarithm)
\begin{eqnarray}
    h_{0}( x ) & = & \log aT + \cdots, ~~~~ x \ll 1        \nonumber   \\
    h_{1}( x ) & \equiv &  0                                    \nonumber  \\
    h_{2}( x ) & = & c_{2} + x^{2} + \cdots, ~~~~ x \ll 1
\end{eqnarray}
      
    Substituting the above equation into Eq.\ref{expand}, we get
\begin{equation}
   \chi_{imp}  =  \log aT + \lambda^{2} T + \lambda^{2} H^{2} + \cdots 
\end{equation}

    In the FL regime, it was shown in Ref.\cite{gogolin2}
\begin{equation}
   \chi_{imp}  = \log aH + \cdots
\label{test3}
\end{equation}

      Actually, the whole crossover functions $ g_{0}(x), h_{0}(x) $ have been
   calculated along the EK line in Ref. \cite{gogolin2}.

\section{ Discussions on experiments and Conclusions}
    In this paper, we brought about the rich phase diagram of a non- magnetic impurity
  hopping between two sites in a metal (Fig.~\ref{picture}). As discussed
  in Sec. IV, the NFL fixed point with the symmetry 2IK(a) is very unlikely
  to be observed, although it has very interesting behaviors $ C_{imp} \sim T,
  \chi^{h}_{imp} \sim \log T $ and $ \sigma(T) \sim 2\sigma_{u}(1- T^{3/2}) $.
  The peculiar behaviors of $ C_{imp}, \chi^{h}_{imp} $ are due to the 'orbital
  field' couples to a {\em non conserved} current. 

  Ralph {\sl et al} found that the experiment data show
  $ \sqrt{T} $ behavior for $ 0.4K< T <T_{K1} \sim 4K $ and concluded
  the two sites system fall in the Quantum Critical (QC) regime 
  controlled by the " 2CK fixed point". They also
  discussed the crossover to the FL regime in the presence of
  magnetic field $ H $ which acts as a channel
  anisotropy of scaling dimension 1/2 \cite{flavor} and in the presence of asymmetry
 of the two sites which acts as a local magnetic field
  of scaling dimension 1/2 \cite{ralph}.
 As first pointed out by MS, even the two sites are exactly symmetric,
  therefore, the two channel are exactly equivalent, there is
 another dimension 1/2 operator $ \Delta_{+} $ which will
 drive the system to FL regime\cite{fisher2}.

  In this paper, we find the " 2CK fixed point " actually is a  line of NFL fixed points 
  which interpolates continuously between the 2CK and the 2IK(a).
  As Gan \cite{gan} showed that under {\em different} canonical transformation
  than employed in this paper \cite{line}, the 2IK model can be mapped to 
  the 2CK model. This paper discussed the two apparent different fixed points
  in a unified framework. Although P-H symmetry breaking is a relevant
   perturbation in the original 2IK model discussed in Refs. \cite{twoimp,gan},
   but its effect is trivial in this model. Because the two models have 
   different {\em global } symmetry, although the fixed point is exactly
   the {\em same}, the allowed boundary operators are {\em different}.

  We discovered a marginal operator in the spin sector which is responsible for this 
  line of NFL fixed points.
  In a real system, there is always P-H symmetry breaking, therefore there is
  always a marginal operator in the charge sector. Eq.\ref{times} show that
   the coefficient of $ \sqrt{T} $ depend on this breaking  any way.  However, 
  the marginal operator identified in this paper is in the spin sector, it, combined with
  the leading irrelevant operator which contributes to the $ \sqrt{T} $ behavior
  of the conductivity, will always generate the dimension $ 1/2 $ relevant operator.
  The existence of the line of NFL fixed points and the existence of the relevant operator
  is closely related.
  There is no reason why the coefficient of this relevant operator is so small
  that it can be neglected in the scaling analysis in the last section.

   The crossover scale from the weak coupling fixed point $ q=0 $
  to {\em any} point on the  line of NFL fixed points  is given by the Kondo scale
  $ T_{K1} \sim D ( \Delta_{K} )^{2} \sim \lambda^{-2} $ ( see Eq.\ref{cond} ),
  the crossover scale
  from a given point on the  line of NFL fixed points  to the corresponding point
  on the  line of FL fixed points  is set by $ \Lambda \sim D (\Delta_{+})^{2} $,
  the finite size
  scaling at finite temperature T leads to the universal scaling function Eq. \ref{cond}.
  Because there is no reliable way to estimate the magnitude of $ \Delta_{+} $
  which is always present. It is very hard to say if the experimental situation 
  do fall in the QC regime controlled by any fixed point on this line.

  The experiment  measured the magnetic field
  dependence of conductance signal. {\em Assuming} $ \Delta_{+} $ is so small that it
  can be neglected, scaling Eqs \ref{test1} in the field $ H $ shows
  that in the FL regime, the conductance should depend on $ \lambda |H| $
  which is consistent with the experimental finding \cite{ralph}.
   The coefficient of $ T^{2} $ of the conductivity should scale as $ H^{-3} $.
   Because $ T_{K} \sim \lambda^{-2} \sim 4K $, the lowest temperature
   achieved in the experiment 
   is $ T_{min} \sim 0.1 K $, if $ 0.1 K < |H| < \lambda^{-1} \sim \sqrt{T_{K}} = 2K $,
   Eq.\ref{test2} shows that the linear $ T $ coefficient of the impurity
   specific heat should scale as $ H^{-2} $. Eq.\ref{test3} shows that the impurity
   susceptibility should scale as $ \log |H| $.
   So far, there is
   no experiment test of these scaling relations in this range of magnetic field.
   It should be possible to extract $ \chi_{imp} $ from experimental data,
   because the impurity does not carry real spin. There is no
  difficulty caused by the conduction electrons and the impurity
  having different Land\'{e} factors \cite{line}. 

  It is very difficult, but not impossible to measure $ \chi^{h}_{imp} $ by adding pressure
  and strain to the system.  Because $ \chi^{h}_{imp} $ here is the hopping susceptibility.
  The difficulty caused by the conduction electrons and the impurity
  having different Land\'{e} factors is not a issue either.
  Unfortunately, {\em no} universal ratios can
  be formed among the amplitudes of the three quantities except at the 2CK point on this
  line of NFL fixed points, this is  because (1) the strain
  couples to a non conserved current (2) this is a line of NFL fixed points  instead of a fixed point.

   The experiment also measured the conductance signal when a finite voltage $ V $ 
   is applied to the point contacts and find  $ e V/ T $ scaling in the temperature range $ 0.4K < T < 4 K $.
   It is not clear how the position on this line of NFL fixed points enter the expression
   of the non-equilibrium conductivity calculations. This question deserve to be addressed
   seriously \cite{delft}. 

  Ref\cite{three} showed that a non-magnetic
  impurity  hopping around 3 sites arranged around a triangular
  is delocalized
 and the metal shows either the one channel Kondo fermi liquid behavior
  or the 2CK  non-fermi liquid behavior. 
 They also conjectured  that there may be
 a " NFL fixed point " possessing local conformal symmetry
  $ SU_{3}(2) \times SU_{2}(3) \times U(1) $
  separating these two fixed points.
  The insights gained from the two sites problem discussed in this paper
  implies that "the NFL fixed point" separating the one channel Kondo
  fixed point and the 2CK fixed point may be a {\em  line of NFL fixed points } instead of
  a NFL fixed point. The symmetry of this  line of NFL fixed points is a interesting
  open question, but it should be {\em smaller}
  than $ SU(3) $, just like the symmetry in the spin sector of the  line of NFL fixed points  in the 
  two sites problem is $ U(1) \times O(1)$ which is smaller than
  $ SU(2) $ \cite{sloppy}. The higher
  symmetry  $SU(3)$ is realized at just one point on this  line of NFL fixed points .
  It was shown in Ref.\cite{three} that the 2CK fixed point with the
  symmetry $SU(2)$ can be realized in $ C_{3v} $ or higher symmetry, because it is
  indeed possible for the ground state of the impurity-electrons complex
  to transform as the $ \Gamma_{3} $ representation of $ C_{3v} $ group,
  therefore a doublet. This NFL fixed point was also shown to be stable. 
  Similarly, Ref \cite{three} pointed out that the stable $ SU(3) $ NFL fixed point
  can be realized in the system of a non-magnetic impurity hopping around the
  tetrahedral or octahedral sites in a cubic crystal when the ground 
  state is a triplet. However, as the symmetry get higher, the NFL fixed
  point with higher symmetry will be more unlikely to be realized by
  experiments, because the number of relevant processes will increase.

\centerline{\bf ACKNOWLEDGMENTS}
   We thank A. W. W. Ludwig, C. Vafa, E. Zaslow for very helpful
   discussions on the spinor
  representation of $ SO(8) $. We also thank D. Fisher, B, Halperin,
  N. Read, S. Sachdev for very interesting discussions. I am indebted
  to A. Millis for very helpful discussions on the last three sections.
   This research was supported by
   NSF Grants Nos. DMR 9106237, DMR9400396 and Johns Hopkins University.

\begin{table} 
\begin{tabular}{ |c|c|c|c|c| }    
 $ O(1) $ & $ U(1) $ & $ O(5) $ & $ \frac{l}{v_{F} \pi}( E-E_{0}) $  & Degeneracy  \\  \hline
  $ R $     &  $  NS_{\delta} $   & $ NS $   &       0                               &    2      \\  \hline
  $  NS $    & $  R_{\delta} $   & $  R $   &       $ \frac{3}{8}-\frac{\delta}{2 \pi} $           &    4      \\ \hline
 $  R $    &  $  NS_{\delta} $   & $ NS+1st $   &  $ \frac{1}{2} $     &    10      \\  \hline
  $  NS $    & $  R_{\delta}+1st $   & $ R $   &    $ \frac{3}{8}+\frac{\delta}{2 \pi} $                    &    8    
\end{tabular} 
\caption{ The finite size spectrum at the  line of NFL fixed points  with the symmetry $ O(1) \times U(1) \times O(5) $
   when $ 0 < \delta < \frac{\pi}{2} $. $ E_{0}=\frac{1}{16} +\frac{1}{2} (\frac{\delta}{\pi})^{2} $. 
  $ NS_{\delta} $ is the state achieved by twisting $ NS $ by the angle $ 2 \delta $,
    namely $ \psi(-l)=-e^{i 2 \delta} \psi(l) $.
  $  R_{\delta} $ is the state achieved by twisting $ R $ by the angle $ 2 \delta $,
    namely $ \psi(-l)=e^{i 2 \delta} \psi(l) $. NS+1st is the first excited state in NS sector {\sl et. al}.
    Only when $ \frac{\pi}{4} < \delta < \frac{\pi}{2} $, the 4th row has lower energy than the 5th row.
    If $ \delta=0 $, the symmetry is enlarged to $ O(1) \times O(7) $, the finite size spectrum of the 2IK fixed point
    is recovered.  If $ \delta=\frac{\pi}{2} $, the symmetry is enlarged to $ O(3) \times O(5) $, the finite size spectrum
   of the 2CK fixed point is recovered. }
\label{nflpositive}
\end{table}

\begin{table} 
\begin{tabular}{ |c|c|c|c|c| }    
 $ O(1) $ & $ U(1) $ & $ O(5)$ & $ \frac{l}{v_{F} \pi}( E-E_{0}) $  & Degeneracy  \\  \hline
  $ R $   &  $  NS_{\delta} $   & $ NS $   &       0                               &    2      \\  \hline
  $  NS $    & $  R_{\delta} $   & $  R $       &       $ \frac{3}{8}+\frac{\delta}{2 \pi} $    &    4      \\ \hline
 $  R $    &  $  NS_{\delta}+1st $   & $ NS $   &  $ \frac{1}{2} +\frac{\delta}{\pi} $     &    4      \\  \hline
 $  R $    &  $  NS_{\delta} $   & $ NS+1st $   &  $ \frac{1}{2} $     &    10      \\  \hline
  $  NS+1st $    & $  R_{\delta}$   & $ R $     & $ \frac{3}{8}+\frac{\delta}{2 \pi} +\frac{1}{2} $         &    4   
\end{tabular} 
\caption{ The finite size spectrum at the  line of NFL fixed points  with the symmetry $ O(1) \times U(1) \times O(5) $
   when $ -\frac{\pi}{2} < \delta < 0 $. 
    Only when $ -\frac{\pi}{4} < \delta < 0 $, the 3rd row has lower energy than the 4th row.
    If $ \delta=0 $, the symmetry is enlarged to $ O(1) \times O(7) $, the finite size spectrum of the 2IK fixed point
    is recovered.  If $ \delta=-\frac{\pi}{2} $, the symmetry is enlarged to $ O(3) \times O(5) $,
    the finite size spectrum of the 2CK fixed point is recovered. }
\label{nflnegative}
\end{table}

\begin{table} 
\begin{tabular}{ |c|c|c|c| }    
 $ O(3) $ & $ O(5) $ &  $ \frac{l}{v_{F}\pi}( E-\frac{3}{16}) $  & Degeneracy  \\  \hline
  $ R $  &   $ NS $   &       0                           &    2      \\  \hline
  $  NS $    &  $  R $       &       $ \frac{1}{8} $    &    4      \\ \hline
 $  R $  &   $ NS+1st $   &  $ \frac{1}{2} $     &    10      \\  \hline
  $  NS+1st $    & $ R $       &       $ \frac{5}{8} $      &    12   \\   \hline
 $  R+1st $  &   $ NS $   &   1      &    6      \\  
 $  R $  &   $ NS+2nd $   &   1      &    20      \\  \hline
 $  NS $  &   $ R+1st $   &  $ 1+\frac{1}{8} $      &    20      \\  
 $  NS+2nd $  &   $ R $   &   $ 1+\frac{1}{8} $      &    12     
\end{tabular} 
\caption{ The finite size spectrum of the 2CK fixed point}
\label{2CK}
\end{table}

\begin{table} 
\begin{tabular}{ |c|c|c|c| }    
 $ O(1) $ & $ O(7) $ &  $ \frac{l}{v_{F} \pi}( E-\frac{1}{16}) $  & Degeneracy  \\  \hline
  $ R $  &   $ NS $   &       0                           &    2      \\  \hline
  $  NS $    &  $  R $       &       $ \frac{3}{8} $    &    8      \\ \hline
 $  R $  &   $ NS+1st $   &  $ \frac{1}{2} $     &    14      \\  \hline
  $  NS+1st $    & $ R $       &       $ \frac{7}{8} $      &    8   \\   \hline
 $  R $  &   $ NS+2nd $   &   1      &    42      \\  
 $  R+1st $  &   $ NS $   &   1      &    2      \\  
\end{tabular} 
\caption{ The finite size spectrum of the 2IK fixed point}
\label{2IK}
\end{table}

\begin{table} 
\begin{tabular}{ |c|c|c|c| }    
  $ U(1) $ & $ O(6) $ & $ \frac{l}{v_{F} \pi}( E-E_{0}) $  & Degeneracy  \\  \hline
  $  NS_{\delta} $   & $ NS $   &       0               &    1      \\  \hline
  $  R_{\delta} $   & $  R $   &       $ \frac{1}{2}-\frac{\delta}{2 \pi} $       &    8      \\ \hline
  $  NS_{\delta} $   & $ NS+1st $   &  $ \frac{1}{2} $     &    6      \\  \hline
  $  R_{\delta}+1st $   & $  R $   &       $ \frac{1}{2}+\frac{\delta}{2 \pi} $       &    16      \\ \hline
  $  NS_{\delta} +1st $   & $ NS $   &  $ \frac{1}{2}+\frac{\delta}{\pi} $     &    2     
\end{tabular} 
\caption{ The finite size spectrum at the  line of FL fixed points  with the symmetry $  U(1) \times O(6) $
   when $ 0 < \delta < \frac{\pi}{2} $. $ E_{0}=\frac{1}{2} (\frac{\delta}{\pi})^{2} $. 
    If $ \delta=0 $, the symmetry is enlarged to $ O(8) $, the finite size spectrum
   of the free fermion fixed point is recovered.
    If $ \delta=\frac{\pi}{2} $, the symmetry is enlarged to $ O(2)\times O(6) $, the finite size spectrum
   of the 2CSFK is recovered.}
\label{flpositive}
\end{table}

\begin{table} 
\begin{tabular}{ |c|c|c|c| }    
  $ U(1) $ & $ O(6) $ & $ \frac{l}{v_{F} \pi}( E-E_{0}) $  & Degeneracy  \\  \hline
  $  NS_{\delta} $   & $ NS $   &       0               &    1      \\  \hline
  $  R_{\delta} $   & $  R $   &       $ \frac{1}{2}+\frac{\delta}{2 \pi} $       &    8      \\ \hline
  $  NS_{\delta}+1st $   & $ NS $   &  $ \frac{1}{2}+\frac{\delta}{\pi} $     &    2      \\  \hline
  $  NS_{\delta} $   & $  NS+1st $   &       $ \frac{1}{2} $       &    6      \\ \hline
  $  NS_{\delta} +2nd $   & $ NS $   &  $ 2(\frac{1}{2}+\frac{\delta}{\pi}) $     &    1      \\  \hline
  $  NS_{\delta} +1st $   & $ NS+1st $   &  $ 1+\frac{\delta}{\pi} $     &    12      \\  \hline
  $  R_{\delta}+1st $   & $  R $   &       $ \frac{3}{2}(1+\frac{\delta}{ \pi}) $       &    16    
\end{tabular} 
\caption{ The finite size spectrum at the  line of FL fixed points  with the symmetry $  U(1) \times O(6) $
   when $ -\frac{\pi}{2} < \delta < 0 $. $ E_{0}=\frac{1}{2} (\frac{\delta}{\pi})^{2} $. 
    If $ \delta=0 $, the symmetry is enlarged to $ O(8) $, the finite size spectrum
   of the free fermion fixed point is recovered.
    If $ \delta=-\frac{\pi}{2} $, the symmetry is enlarged to $ O(2)\times O(6) $, the finite size spectrum
   of the 2CSFK is recovered.}
\label{flnegative}
\end{table}

\begin{table} 
\begin{tabular}{ |c|c|c| }    
 $ O(8) $  &  $ \frac{l}{v_{F} \pi} E $  & Degeneracy  \\  \hline
   $ NS $   &       0                           &    1      \\  \hline
   $  R $     &       $ \frac{1}{2} $    &    16      \\ 
  $ NS+1st $   &  $ \frac{1}{2} $     &    8      \\  \hline
  $  NS+2nd $    &       1      &    28   \\   \hline
  $ NS+3rd $   &  $ \frac{3}{2} $      &    64      \\  
 $  R+1st $  &     $  \frac{3}{2} $      &    8        
\end{tabular} 
\caption{ The finite size spectrum of the free fermions with both NS and R sectors \cite{double1}}
\label{free}
\end{table}

\begin{table} 
\begin{tabular}{ |c|c|c|c| }    
 $ O(2) $ & $ O(6) $ &  $ \frac{l}{v_{F} \pi}( E-\frac{1}{8}) $  & Degeneracy  \\  \hline
  $ R $  &   $ NS $   &       0                           &    2      \\  \hline
  $  NS $    &  $  R $       &       $ \frac{1}{4} $    &    8      \\ \hline
 $  R $  &   $ NS+1st $   &  $ \frac{1}{2} $     &    12      \\  \hline
  $  NS+1st $    & $ R $       &       $ \frac{3}{4} $      &    16   \\   \hline
 $  R $  &   $ NS+2nd $   &   1      &    30      \\  
 $  R+1st $  &   $ NS $   &   1      &    4      \\  
\end{tabular} 
\caption{ The finite size spectrum of the 2CSFK fixed point \cite{double2}}
\label{2CSFK}
\end{table}

\appendix

\section{ The finite size spectrum of one complex fermions }

   In this appendix, we closely follow the notations of \cite{ginsbarg}.

  The energy momentum tensor of one complex fermion in the complex plane is:
\begin{equation}
    T( z) =: \psi^{*}(z) \partial \psi(z) :
\end{equation}
   Where the complex fermion can be written in terms of two Majorana fermion:
\begin{equation}
  \psi(z) = \frac{1}{\sqrt{2}} (\psi_{1}(z) +i \psi_{2} (z) ), ~~~~
  \psi^{*}(z) = \frac{1}{\sqrt{2}} (\psi_{1}(z) -i \psi_{2} (z) )
\end{equation}

    The mode expansions of the complex fermion are 
\begin{eqnarray}
   i \psi(z) & = & \sum_{n} \psi_{n} z^{-n-1/2},~~~ \psi_{n}=\frac{1}{\sqrt{2}} (\psi^{1}_{n}+i \psi^{2}_{n} ) 
                       \nonumber  \\
   i \psi(z) & =  & \sum_{n} \psi^{*}_{n} z^{-n-1/2}, ~~~ \psi_{n}=\frac{1}{\sqrt{2}} (\psi^{1}_{n}-i \psi^{2}_{n} ) 
\end{eqnarray}

    The modes satisfy the commutation relations:
\begin{equation}
   \{\psi_{n},\psi_{m} \}= \{\psi^{*}_{n},\psi^{*}_{m}\}=0,~~~~\{\psi_{n},\psi^{*}_{m}\}=\delta_{n+m,0}
\end{equation}

   Under the conformal transformation $ z=e^{w} $,
   mapping the cylinder, labeled by $ w=\tau + i \sigma $, to the plane,
 labeled by  $ z $
\begin{equation}
 \psi(w)= \sum_{n} \psi_{n} e^{-n w},~~~~ \psi^{*}(w)= \sum_{n} \psi^{*}_{n} e^{-n w}
\end{equation}
  Where $ n \in Z-\theta $ in order to satisfy the boundary conditions on the cylinder
\begin{equation}
   \psi( w+ i 2 \pi )= e^{i 2 \pi \theta } \psi(w),~~~ 0< \theta <1
\end{equation}

   The energy of the complex fermion on the cylinder is
\begin{equation}
  H=L_{0}= \sum_{ n \in Z } (n-\theta) : \psi^{*}_{n-\theta} \psi_{-n+\theta} :
\end{equation}
 
    From the commutation relations
\begin{equation}
  [ L_{0}, \psi_{n-\theta} ]=-(n-\theta)\psi_{n-\theta},~~~~
  [ L_{0}, \psi^{*}_{n-\theta} ]=-(n-\theta)\psi^{*}_{n-\theta}
\end{equation}
   
   and the fact that the regularity at $ \tau= -\infty $ requires the positive modes annihilate the vacuum
\begin{equation}
  \psi_{n-\theta} |0 \rangle_{\theta}=0,~~~~ n-\theta >0 
\end{equation}

    We can construct the Hilbert space
\begin{equation}
  \cdots \psi^{*}_{-m-\theta} \psi_{-n-\theta}|0 \rangle_{\theta}
\label{hilbert}
\end{equation}
   Where $ m,n $ are no negative integers.
   
   The energy of the twisted vacuum $ |0 \rangle_{\theta} $ is
\begin{equation}
   E_{0}=f(\theta)=-\frac{1}{24} +\frac{1}{2} ( \frac{\delta}{\pi} )^{2},~~~~\theta= \frac{1}{2} + \frac{\delta}{\pi}
\end{equation}

   The energies of the excited states in Eq.\ref{hilbert} are
\begin{equation}
   E= E_{0} + m+\theta+n+\theta + \cdots
\end{equation}

  If  $ \theta= 0 $, then the states are in Ramond ( periodic ) sector, because of zero modes, the ground
 state degeneracy is 4, the ground states are $ |0 \rangle, \psi_{0} |0 \rangle, \psi^{*}_{0} |0 \rangle,
  \psi_{0} \psi^{*}_{0} |0 \rangle $. If $ \theta=\frac{1}{2} $, the states are in Neveu-Schwarz (anti-periodic)
  sector.

\section{ The derivation of the boundary conditions }

  $ SO(8) $ has three 8-dimensional representations, one vector representation
 $ 8_{v} $; two spinor representations, one $ 8_{c} $ with positive chirality $ \Gamma=1 $,
 another $ 8_{s}$ with negative chirality $ \Gamma=-1 $. They can be bosonized as
\begin{equation}
 \psi_{\mu}= \left( \begin{array}{c} e^{-i \Phi_{1,\uparrow}}   \\
                         e^{-i \Phi_{1,\downarrow} }   \\
                         e^{-i \Phi_{2,\uparrow} }   \\
                         e^{-i \Phi_{2,\downarrow} }  \end{array}  \right ),~~~~
 C_{\mu}= \left( \begin{array}{c} e^{-i \Phi_{c}}   \\
                         e^{-i \Phi_{s} }   \\
                         e^{-i \Phi_{f} }   \\
                         e^{-i \Phi_{sf} }  \end{array}  \right ),~~~~
 S_{\mu}= \left( \begin{array}{c} e^{-i \Phi_{c,s}}   \\
                         e^{-i \Phi_{s,s} }   \\
                         e^{-i \Phi_{f,s} }   \\
                         e^{-i \Phi_{sf,s} }  \end{array}  \right )
\label{vcs}
\end{equation}
  Where $\Phi_{c},\Phi_{s},\Phi_{f},\Phi_{sf} $ are defined by {\em first}
  triality transformation Eq.\ref{second}.

  $ \Phi_{c,s},\Phi_{s,s},\Phi_{f,s},\Phi_{sf,s} $ are defined by {\em second}
   "triality transformations"
\begin{equation}
   \left ( \begin{array}{c} \Phi_{c,s} \\
 \Phi_{s,s}  \\
 \Phi_{f,s}  \\
 \Phi_{sf,s} \end{array}  \right )
  =\frac{1}{2} \left ( \begin{array}{c} \Phi_{1 \uparrow }- \Phi_{1\downarrow }+
            \Phi_{2 \uparrow }+ \Phi_{2 \downarrow }  \\
           \Phi_{1 \uparrow }+ \Phi_{1\downarrow }+
            \Phi_{2 \uparrow }- \Phi_{2 \downarrow }  \\
           \Phi_{1 \uparrow }- \Phi_{1\downarrow }-
            \Phi_{2 \uparrow }- \Phi_{2 \downarrow }  \\
           \Phi_{1 \uparrow }+ \Phi_{1\downarrow }-
            \Phi_{2 \uparrow }+ \Phi_{2 \downarrow }  \end{array} \right )
\label{tri}
\end{equation}

   Note the only difference between $ C_{\mu} $ and $ S_{\mu} $ is the change of sign
   of  $\Phi_{1\downarrow} $.

     In the  basis of Eq.\ref{second}, we can rewrite Eq.\ref{vcs} as
\begin{equation}
 C_{\mu}= \left( \begin{array}{c} e^{-i \Phi_{c}}   \\
                         e^{-i \Phi_{s} }   \\
                         e^{-i \Phi_{f} }   \\
                         e^{-i \Phi_{sf} }  \end{array}  \right ),~~~~
 \psi_{\mu}= \left( \begin{array}{c} e^{-i \Phi_{1,\uparrow}}   \\
                         e^{-i \Phi_{1,\downarrow} }   \\
                         e^{-i \Phi_{2,\uparrow} }   \\
                         e^{-i \Phi_{2,\downarrow} }  \end{array}  \right ),~~~~
 S_{\mu}= \left( \begin{array}{c} e^{-i \Phi_{c,s}}   \\
                         e^{-i \Phi_{s,s} }   \\
                         e^{-i \Phi_{f,s} }   \\
                         e^{-i \Phi_{sf,s} }  \end{array}  \right )
\end{equation}

    The other two sets of bosons are expressed in terms of those in Eq.\ref{second}
\begin{equation}
   \left ( \begin{array}{c} \Phi_{1,\uparrow} \\
 \Phi_{1,\downarrow}  \\
 \Phi_{2,\uparrow}  \\
 \Phi_{2,\downarrow} \end{array}  \right )
  =\frac{1}{2} \left ( \begin{array}{c} \Phi_{c }+ \Phi_{s }+ \Phi_{f }+ \Phi_{sf }  \\
  \Phi_{c }- \Phi_{s }+ \Phi_{f }- \Phi_{sf }      \\
  \Phi_{c }+ \Phi_{s }- \Phi_{f }- \Phi_{sf }  \\
  \Phi_{c }- \Phi_{s }- \Phi_{f }+ \Phi_{sf }  \end{array}   \right ),~~~~  
   \left ( \begin{array}{c} \Phi_{c,s} \\
 \Phi_{s,s}  \\
 \Phi_{f,s}  \\
 \Phi_{sf,s} \end{array}  \right )
  =\frac{1}{2} \left ( \begin{array}{c} \Phi_{c }- \Phi_{s }+ \Phi_{f }+ \Phi_{sf }  \\
  \Phi_{c }+ \Phi_{s }+ \Phi_{f }- \Phi_{sf }      \\
  \Phi_{c }- \Phi_{s }- \Phi_{f }- \Phi_{sf }  \\
  \Phi_{c }+ \Phi_{s }- \Phi_{f }+ \Phi_{sf }  \end{array}   \right )  
\label{basis}
\end{equation}

   It is evident that in the first basis of Eq.\ref{second}, $ \psi_{\mu} $ transforms as $ 8_{c} $
 and $ C_{\mu} $ transforms as $ 8_{v} $, namely, $\psi_{\mu} $ and $ C_{\mu} $
 exchange roles. It is important to observe that the only difference between
 $\psi_{\mu} $ ( $\Gamma=1$) and $ S_{\nu} $( $\Gamma=-1 $ ) is the change of sign
 of $\Phi_{s} $.

   Similarly, in the basis of Eq.\ref{tri}, 
   $ \psi_{\nu} $ transforms as $ 8_{s} $ and $ S_{\nu} $ transforms
   as $ 8_{v} $, namely, $\psi_{\nu} $ and $ S_{\nu} $ exchange roles.

   The complex fermions in Eq.\ref{ek} are $ C_{\nu} $ fermions.

\subsection{ The derivation by bosonization}

   In Eq.\ref{jinwu}, we bosonize {\em new} sets of complex fermions in terms of {\em new}
   sets of bosons $ \Phi^{n}_{s}, \Phi^{n}_{sf} $.

    Replacing $\Phi_{s},\Phi_{sf} $ by $\Phi^{n}_{s},\Phi^{n}_{sf} $
    in Eq.\ref{basis}, we can construct $\psi^{n}_{\nu}, S^{n}_{\nu} $.

    Eq.\ref{obi} and \ref{basis} lead to 
\begin{equation}
 \psi^{n}_{i \alpha,L }  =   e^{  \pm i \theta/2 } S^{n}_{i \bar{\alpha},R}, ~~~~~
 S^{n}_{i \alpha,L }  =   e^{  \pm i \theta/2 } \psi^{n}_{i \bar{\alpha},R}
\label{yin}
\end{equation}
  where in the exponential, we take $ + $, if $\alpha=\uparrow $ and $-$, if $\alpha=\downarrow $.

  Because  the basis \ref{jinwu} is related to the basis \ref{ek} by a $ SO(8) $ rotation
  matrix in  $ 8_{v} $, therefore $ \psi^{n}_{i, \alpha}, S^{n}_{i, \alpha} $ are
  related to $ \psi_{i, \alpha}, S_{i, \alpha} $ by the two rotations in $ 8_{c}, 8_{s} $ respectively.

   We can determine the two rotations in $ 8_{c}, 8_{s} $ by the following two steps.

 (1) Under $\Phi_{c} \rightarrow \Phi_{c} + \lambda, \Phi_{f} \rightarrow \Phi_{f} + \lambda $,
      $ \lambda $ is an arbitrary angle.

       Eq.\ref{basis} dictates
\begin{eqnarray}
    \psi^{\prime}_{i \uparrow}= c_{11} \psi_{i,\uparrow} + c_{12} \psi_{i,\downarrow}  \nonumber  \\
    \psi^{\prime}_{i \downarrow}= c_{21} \psi_{i,\uparrow} + c_{22} \psi_{i,\downarrow} 
\end{eqnarray}
 
 (2) Under $ \Phi_{s} \rightarrow \Phi_{s}+ \pi $, $ a_{s} \rightarrow -a_{s}, b_{s} \rightarrow -b_{s} $
    therefore $ \Phi^{n}_{s} \rightarrow -\Phi^{n}_{s}+\pi,   
    \Phi^{n}_{sf} \rightarrow -\Phi^{n}_{sf}  $.
    Under $ \Phi_{sf} \rightarrow \Phi_{sf}+ \pi $, $ a_{sf} \rightarrow -a_{sf}, b_{sf} \rightarrow -b_{sf} $
    therefore $ \Phi^{n}_{s} \rightarrow -\Phi^{n}_{s},   
    \Phi^{n}_{sf} \rightarrow -\Phi^{n}_{sf} + \pi  $.

       Eq.\ref{basis} dictates
\begin{eqnarray}
 \psi^{n}_{i \uparrow }   = \psi_{i,+}=  \frac{1}{\sqrt{2}}
      ( \psi_{i \uparrow} +  \psi_{i \downarrow}  )   \nonumber   \\
 \psi^{n}_{i \downarrow }   = \psi_{i,-}=  \frac{1}{\sqrt{2}}
      ( \psi_{i \uparrow} -  \psi_{i \downarrow}  )
\label{finish1}
\end{eqnarray}

   Remember $ S $ differ from $ \psi $ by the change of sign of $ \Psi_{s} $, we get
   the similar relations for $ S $ spinor
\begin{eqnarray}
 S^{n}_{i \downarrow }  = S_{i,+}=  \frac{1}{\sqrt{2}}
      ( S_{i \uparrow} +  S_{i \downarrow}  )   \nonumber   \\
 S^{n}_{i \uparrow }  = S_{i,-}=  \frac{1}{\sqrt{2}}
      ( S_{i \uparrow} -  S_{i \downarrow}  )
\label{finish2}
\end{eqnarray}

  Eqs.\ref{yin},\ref{finish1},\ref{finish2} give the boundary conditions at the  line of NFL fixed points 
  Eq.\ref{spinor1}.

\subsection{ The derivation by $ \gamma $ matrix }

    In this subsection, we follow Ref.\cite{witten}, but use different conventions.

   $ J_{kl} $ are the  28 generators of $ SO(8) $ and 4 commuting generators $ W_{k} $
   form a Cartan subalgebra of $ SO(8) $
\begin{eqnarray}
   (J_{kl})_{mp} & = & \delta_{km} \delta_{lp}-\delta_{kp} \delta_{lm},~~~k,l=1,\cdots,8  \nonumber  \\
     W_{k} & = &  J_{2k-1,2k},~~~~ k=1,2,3,4.
\end{eqnarray}

  The eight Majorana fermions in Eq.\ref{ek} transform as $ 8_{v} $. Eq.\ref{vectorf} can be written
  as $ e^{\theta J_{38}} $ which is a matrix in $ SO(8) $ (therefore in $ 8_{v} $ ).

   $\psi_{\nu}, S_{\nu} $ will transform as $ 16 \times 16 $ spinor representation of $ SO(8) $.
  The $ 16 \times 16 $ matrices can be written in the block form
\begin{equation}
   \Gamma^{i}=  \left( \begin{array}{cc}
        0    &   \gamma^{i}   \\
       \gamma^{i \prime}   &   0   \\
       \end{array}  \right ),~~~~i=1,\cdots,8
\end{equation}

      $ \Gamma^{i} $ satisfy $ \{ \Gamma^{i},\Gamma^{j} \}=2 \delta^{ij} $ if
\begin{eqnarray}
    \gamma^{i} \gamma^{j \prime} + \gamma^{j} \gamma^{i \prime} =2 \delta^{ij} \nonumber  \\
    \gamma^{i \prime} \gamma^{j}  +\gamma^{j \prime} \gamma^{i} =2 \delta^{ij}  
\end{eqnarray}

    The generators of the $ 16 \times 16 $ representation are
\begin{equation}
   \Gamma^{ij}= \frac{1}{4} [\Gamma^{i},\Gamma^{j}]=
   \frac{1}{4}  \left( \begin{array}{cc}
        \gamma^{i} \gamma^{j \prime} -\gamma^{j} \gamma^{i \prime}    &   0   \\
         0 &  \gamma^{i \prime} \gamma^{j}  -\gamma^{j \prime} \gamma^{i}     \\
       \end{array}  \right ) = \left( \begin{array}{cc}
         \gamma^{i j}_{c}    &   0   \\
         0 &  \gamma^{i j}_{s}  \\ \end{array}  \right )
\end{equation}
  Where $ \gamma^{ij}_{c}=\frac{1}{2} \gamma^{i} \gamma^{j \prime},
  \gamma^{ij}_{s}=\frac{1}{2} \gamma^{i \prime} \gamma^{j} $ are the generators
  of $ 8_{c} $ and $ 8_{s} $ respectively.

  Under $ \Phi_{c} \rightarrow \Phi_{c} + \theta $, Eq.\ref{ek} transform under $ e^{ \theta J_{12} } $,
  from Eq.\ref{basis}, $\psi_{\mu}, S_{\mu} $ both transform under
  $ e^{ \theta/2 ( J_{12}+ J_{34} +J_{56} + J_{78} )} $.
  Under $ \Phi_{s} \rightarrow \Phi_{s} + \theta $, Eq.\ref{ek} transform under $ e^{ \theta J_{34} } $,
  from Eq.\ref{basis}, $\psi_{\mu}, S_{\mu} $ transform under
  $ e^{ \theta/2 (J_{12}- J_{34} +J_{56} - J_{78})} $ and
  $ e^{ -\theta/2 (J_{12}- J_{34} +J_{56} - J_{78})} $ respectively.
  Under $ \Phi_{f} \rightarrow \Phi_{f} + \theta $, Eq.\ref{ek} transform under $ e^{ \theta J_{56} } $,
  from Eq.\ref{basis}, $\psi_{\mu}, S_{\mu} $ both transform under
  $ e^{ \theta/2 ( J_{12}+ J_{34} -J_{56} - J_{78} )} $.
  Under $ \Phi_{sf} \rightarrow \Phi_{sf} + \theta $, Eq.\ref{ek} transform under $ e^{ \theta J_{78} } $,
  from Eq.\ref{basis}, $\psi_{\mu}, S_{\mu} $ both  transform under
  $ e^{ \theta/2 ( J_{12}+ J_{34} -J_{56} - J_{78} )} $. Therefore, omitting
  subscript $ c $, we conclude
\begin{eqnarray}
 \gamma_{12} & = & \frac{1}{2}( J_{12}+J_{34}+J_{56}+J_{78} )=\gamma_{12,s}  \nonumber  \\ 
 \gamma_{34} & = &  \frac{1}{2}( J_{12}-J_{34}+J_{56}-J_{78} )=-\gamma_{34,s}   \nonumber  \\
 \gamma_{56} & = & \frac{1}{2}( J_{12}+J_{34}-J_{56}-J_{78} )=\gamma_{56,s}      \nonumber  \\
 \gamma_{78} & = & \frac{1}{2}( J_{12}-J_{34}-J_{56}+J_{78} )=\gamma_{78,s} 
\label{phis}
\end{eqnarray}

  The 8 $ \gamma_{i} $ matrices can be expressed as the direct products of $ 2 \times 2 $ blocks
\begin{eqnarray}
 \gamma_{1} & = & \epsilon \times 1 \times \sigma_{3},~~~~ \gamma_{2}=\epsilon \times 1 \times \sigma_{1}  \nonumber  \\
 \gamma_{3} & = & 1 \times \sigma_{3} \times \epsilon,~~~~ \gamma_{4}=1 \times 1 \times 1  \nonumber  \\
 \gamma_{5} & = & \epsilon \times \epsilon \times \epsilon,
     ~~~~ \gamma_{6}=\sigma_{1} \times \epsilon \times 1  \nonumber  \\
 \gamma_{7} & = & 1 \times \sigma_{1} \times \epsilon,~~~~ \gamma_{8}=\sigma_{3} \times \epsilon \times 1
\label{pauli}
\end{eqnarray}
   Where  $ \epsilon= i\sigma_{2} $ and $ \sigma_{i} $ are the three Pauli matrices.

   It is easy to see $ \gamma_{4} $ is an identity matrix, the other 7 matrices are real and anti-symmetric.

   It can be checked that indeed 
   $ \gamma^{ij}_{c}=\frac{1}{2} \gamma^{i} \gamma^{j \prime},
  \gamma^{ij}_{s}=\frac{1}{2} \gamma^{i \prime} \gamma^{j} $ satisfy Eq.\ref{phis}.

   If Eq.\ref{ek} transform under $ e^{ \theta J_{38} } $, then $ \psi_{\mu} $ and $ S_{\mu} $
   transform under $ e^{\theta \gamma_{38}} $ and $ e^{\theta \gamma_{38,s}} $ respectively.

  Using Eq.\ref{pauli}, straightforward calculation show that $ e^{ \theta \gamma_{38 } } $
  give Eq.\ref{last}.

   Under the first equation in Eq.\ref{vectornf}, $ \Phi_{s} \rightarrow -\Phi_{s} $, from Eq.\ref{basis},
 it is easy to see $ \psi_{\mu} $ and $ S_{\mu} $ transform to each other, therefore lead to
  Eq.\ref{spinor1}. 

\section{ The calculations in the {\em old} boson basis at the
              additional NFL fixed point} 

  {\em In contrast to} the boundary conditions Eq.\ref{vectornf} of the  line of NFL fixed points ,
 The boundary conditions Eq.\ref{twofold} are in one of the four Cartan subalgebra,
 therefore can be expressed in terms of chiral bosons in Eq.\ref{second}:
\begin{equation}
\Phi_{s,L}(0) = - \Phi_{s,R}(0) + \pi
\end{equation}

   In terms of physical fermions, it reads:
\begin{equation}
\psi_{ i \alpha,L}(0) = e^{i \pi j_{\alpha}}  S_{i \alpha,R}(0)
\label{spinor2}
\end{equation}
    where $ j_{\alpha} =\pm \frac{1}{2} $.

 The five operators in Eq.\ref{twofold1} can be written in terms of
 the bosons introduced in Eq.\ref{second},
\begin{eqnarray}
    \frac{\partial}{\partial \tau} \cos \Phi_{sf}(0,\tau),~~~ 
    \frac{\partial}{\partial \tau} \cos \Phi_{s}(0,\tau)    \nonumber   \\
    \cos 2\Phi_{sf}(0,\tau) -\frac{1}{2} :( \partial \Phi_{sf}(0,\tau))^{2}:  \nonumber  \\
    -\cos 2\Phi_{s}(0,\tau) -\frac{1}{2} :( \partial \Phi_{s}(0,\tau))^{2}:  \nonumber  \\
   \gamma_{2}( \cos 2\Phi_{s}(0,\tau) -\frac{1}{2} :( \partial \Phi_{s}(0,\tau))^{2}:)
\end{eqnarray}

  The four dimension 5/2 operators are
\begin{eqnarray}
   ( :\cos 2\Phi_{s}(0,\tau): -\frac{1}{2} :( \partial \Phi_{s}(0,\tau))^{2}:) \cos \Phi_{sf}(0,\tau) 
                               \nonumber  \\
   ( :\cos 2\Phi_{s}(0,\tau): -\frac{1}{2} :( \partial \Phi_{s}(0,\tau))^{2}:) \cos \Phi_{s}(0,\tau) 
                                \nonumber \\
    \frac{\partial^{2}}{\partial \tau^{2}} \cos \Phi_{sf}(0,\tau),~~~  
   \frac{\partial^{2}}{\partial \tau^{2}} \cos \Phi_{s}(0,\tau) ~~~~~~
\end{eqnarray}

      Very similar arguments to those in Sec. IV lead to $ \sigma(T) \sim 2 \sigma_{u}
    (1+  T^{3/2}  ) $.


\begin{references}
\bibitem{atten}  {\em In contrast to} the 2CK, we {\em don't }
 reverse the spin here, so this Time Reversal symmetry is {\em different }
 than that in the 2CK.
\bibitem{anti} If two operators anticommute with each other, they will not generate
               any new terms.
\bibitem{scaling} Because $ H= U^{-1} H^{\prime} U $, the actual scaling field in $ H^{\prime} $
      is $ \psi^{\prime}_{s}(x) $ instead of $ \psi_{s,i}(x) $. In the following,
      when  deriving the leading or subleading irrelevant operators in $ H^{\prime} $, we will
      replace $ \psi_{s,i}(x) $ directly by $ \psi^{\prime}_{s}(x) $. Then Eq.\ref{general}
      will yield the corresponding operators in $ H $.
\bibitem{known} if keeping $\gamma_{1}=\gamma_{2}=1 $, we get the weak coupling
                RG flow Eq.\ref{weak1}.
\bibitem{des} See \cite{flavor} for very similar discussions on the operator contents.
                Similar operators were also identified near the 2IK fixed point discussed
                in Ref.\cite{twoimp}.
\bibitem{cut} For simplicity, we omit the cocycle factors in the following expressions.
\bibitem{check} We can also check the operator contents in the two special
    cases directly: If $ \Delta_{-}=0 $, namely at the 2CK fixed point, $ \widehat{a} \sim b_{sf} $,
    the spin-0 operator is $ q b_{sf} a_{s,i} b_{s,i} $;
    If $ \Delta_{1}=0 $, namely at the 2IK fixed point, $ \widehat{a} \sim b_{s,i} $,
    the spin-0 operator is $ y b_{s,i} a_{s,i} b_{sf} $.
\bibitem{ambi} There is a sign ambiguity due to the spinor nature of the
    representation, but, if $ \theta=\pi $, the fixed point is the 2CK, so all the
    2 channel and 2 parity should make the same contribution to the conductivity,
    this fix the sign of Eq.\ref{sign2}.
\bibitem{omit} In the following, we don't write out the other trivial factors.
\bibitem{double1} In NRG, either NS or R sector is chosen.
\bibitem{double2}  This finite size spectrum actually has $ g=2 $. If we reduce
  the degeneracy in column 4 by factor 2, we get the finite size spectrum
  of $ g=1 $, but still with the symmetry $ O(2) \times O(6) $. See Ref.\cite{spinflavor}
  for the breaking of the degeneracy from $ g=2 $ to $ g=1 $.
\bibitem{sloppy} We are sloppy about the discrete symmetry $Z_{2}$, however,
     the discrete symmetry may be important. For example, $ O(2) $ is the
     semi-product of $ U(1) $ and $ Z_{2} $, their difference is clearly
     shown along the  line of FL fixed points  in Fig.\ref{picture}.
\end{references}
\end{document}